\documentclass[sigconf,screen]{acmart}

\usepackage{balance}

\AtBeginDocument{%
  }

\usepackage[inline]{enumitem}
\usepackage{multirow}

\copyrightyear{2025}
\acmYear{2025}
\setcopyright{cc}
\setcctype{by}
\acmConference[FAccT '25]{The 2025 ACM Conference on Fairness, Accountability, and Transparency}{June 23--26, 2025}{Athens, Greece}
\acmBooktitle{The 2025 ACM Conference on Fairness, Accountability, and Transparency (FAccT '25), June 23--26, 2025, Athens, Greece}\acmDOI{10.1145/3715275.3732139}
\acmISBN{979-8-4007-1482-5/2025/06}

\begin{document}

\title{Evaluating the Contextual Integrity of False Positives in Algorithmic Travel Surveillance}

\author{Alina Wernick}
\affiliation{%
  \institution{CZS Institute for Artificial Intelligence and Law \\
University of T\"ubingen}
  \city{T\"ubingen}
  \country{Germany}}
\email{alina.wernick@uni-tuebingen.de}

\author{Alan Medlar}
\affiliation{%
  \institution{Department of Computer Science \\
University of Helsinki}
  \city{Helsinki}
  \country{Finland}}
\email{alan.j.medlar@helsinki.fi}

\author{Sofia S\"oderholm}
\affiliation{%
  \institution{Legal Tech Lab, Faculty of Law \\
University of Helsinki}
  \city{Helsinki}
  \country{Finland}}
\email{sofia.soderholm@helsinki.fi}

\author{Dorota G{\l}owacka}
\affiliation{%
  \institution{Department of Computer Science \\
University of Helsinki}
  \city{Helsinki}
  \country{Finland}}
\email{dorota.glowacka@helsinki.fi}


\begin{abstract}
International air travel is highly surveilled. While surveillance is deemed necessary for law enforcement to prevent and detect terrorism and other serious crimes, even the most accurate algorithmic mass surveillance systems produce high numbers of false positives. Despite the potential impact of false positives on the fundamental rights of millions of passengers, algorithmic travel surveillance is lawful in the EU. However, as the system's processing practices and accuracy are kept secret by law, it is unknown to what degree passengers are accepting of the system’s interference with their rights to privacy and data protection. 

We conducted a nationally representative survey of the adult population of Finland (N=1550) to assess their attitudes towards algorithmic mass surveillance in air travel and its potential expansion to other travel contexts. Furthermore, we developed a novel approach for estimating the threshold, beyond which, the number of false positives breaches individuals' perception of contextual integrity. Surprisingly, when faced with a trade-off between privacy and security, even very high false positive counts were perceived as legitimate. This result could be attributed to Finland's high-trust cultural context, but also raises questions about people’s capacity to account for privacy harms that happen to other people. We conclude by discussing how legal and ethical approaches to legitimising algorithmic surveillance based on individual rights may overlook the statistical or systemic properties of mass surveillance.
\end{abstract}




\maketitle

\section{Introduction}

\subsection{Algorithmic Mass Surveillance}
“From initial transaction to exiting the airport at the final destination” air travel is governed by code that determines everything from check-in to boarding the plane \cite{kitchin2014code}. Air travel is also highly surveilled, with passengers subject to complex, digital mass surveillant assemblages by various agencies, such as air travel companies, border control and law enforcement \cite{haggerty2017surveillant}. Often, these processes of surveillance are invisible to passengers \cite{kitchin2014code}. In this article, we focus on a specific mass surveillance practice adopted in air travel - the algorithmic processing of passenger name records (PNR). The PNR system originated in the US in the aftermath of the September 11, 2001 terror attacks to monitor terrorist threats \cite{kitchin2014code}, but was gradually adopted elsewhere, including in the EU where it is mandated by the Passenger Name Record Directive (PNRD) \cite{PNRD} that established the conditions for the processing of PNR data for the prevention, detection, investigation and prosecution of terrorist offences and serious crimes \cite{olsen2021beyond}.

 PNR analysis is a form of predictive policing: the detection of potential offenders is performed automatically against databases and using pre-determined criteria that can be updated on the basis of further analysis of PNR data \cite{Korff2015}. 
 As such, the PNR system ``triages'' which passengers should be subject to follow-up scrutiny by competent authorities \cite{palmiotto2024decision}. 
 The use of AI for predictive policing may expose its subjects to discriminatory biases \cite{lum2016predict, jefferson2018predictable, williams2019data, trials2021automating, browning2021stop} and the potential discriminatory impact of the algorithmic processing of PNR data has been subject both to legal debate \cite{Korff2015, korff2021opinion, korff2023did, olsen2021beyond, brouwer2023future, haitsma2023regulating} and judicial scrutiny in the CJEU decision \textit{Ligue des Droits Humains} \cite{CJEUCase}. 
 However, in this paper, we study the PNR system as a form of mass surveillance that interferes with the fundamental rights to privacy and data protection of 
 millions of EU citizens (Arts. 7 and 8 of the Charter of Fundamental Rights of European Union \cite{CFREU}) \cite{CJEUCase, korff2021opinion}.  The PNR system is subject to the base rate fallacy \cite{bar1980base, EDri}, a problem unaddressed by the CJEU \cite{korff2023did, Thönnes, thonnes2022cautious},
 that arises from the exceptionally low prevalence of terrorists and perpetrators of other serious crimes in the general population.
 As a result, even a surveillance system with a very low false positive rate will flag far more innocent passengers (false positives) than actual terrorists (true positives) as potential criminal threats \cite{munk2017100, korff2023did, EDri}. Indeed, assuming an accuracy of 99.9 percent, the PNR system is expected to produce 500,000 false positives per year within the EU \cite{EDri, korff2023did}. Any efforts to debias the system would not remove this statistical property of the model, 
 exposing a large number of passengers flagged as false positives to further scrutiny by the authorities.

The PNR system is also subject to multiple layers of opacity, making it difficult to observe and interfere with. Unlike physical surveillance methods, such as closed-circuit television cameras, passengers are generally unaware of the processing of their PNR data. Indeed, the methods, procedures and criteria for the automated processing of personal data are opaque \cite{ulbricht2018big} and can be kept secret on the grounds of law enforcement and national security (Art. 15, Law Enforcement Directive (LED) \cite{LED}). Furthermore, when the initial assessment is conducted using AI, the grounds for investigating and potentially stopping an individual passenger can be unexplainable, interfering with the passenger's right to an effective judicial remedy (Art. 47 CFREU, \cite{CJEUCase, korff2023did}). Moreover, the PNR system is vulnerable to function creep, where surveillance methods adopted in one context gradually and imperceptibly expand into new domains or are used to serve other purposes \cite{wood2006report, koops2021concept, tzanou2010eu}. For example, from specific flights to all flights or from air travel to sea travel.

\subsection{Legitimisation Approaches}
Globally, the EU is known for a strict, rights- and risk-based approach to technology regulation \cite{gellert2016we, de2022european}, reflected in the General Data Protection Regulation (GDPR) \cite{GDPR} and the Artificial Intelligence Act (AI act) \cite{AIA}. In the rights-based approach, the normative values at stake are fundamental rights, that influence legislative activity and legal interpretation in the EU in a top-down manner. Fundamental rights are not absolute, but can be limited on the basis of the principle of proportionality and on the grounds of public security \cite{CFREU, CJEUCase}. The CJEU ruled that the PNRD's exceptions to fundamental rights must be interpreted restrictively, and that surveillance shall not apply indiscriminately to all intra-EU flights. It also stressed that automated assessment of PNR data may, generally, not be carried out using self-learning systems \cite{CJEUCase}. The AI Act renders law enforcement applications of AI as high-risk because they have the potential to perpetuate discrimination if their training data reflects historical biases and other societal inequalities (AI Act, Rec. 59). However, as discussed in the following section, the PNRD's and AI Act's limits to the use of self-learning PNR systems are not absolute (AI Act Arts. 3 (1), 8, Annex III, Art. 6 (e)) \cite{thonnes2022directive}. 

However, even though the PNR directive requires that assessment criteria be “defined in a manner which keeps to a minimum the number of innocent people wrongly identified by the system” (PNRD, Rec 7), neither the directive nor the CJEU addresses the problem of the base rate fallacy, 
 exposing millions of innocent citizens to the possibility of being scrutinised by the authorities
\cite{korff2021opinion, korff2023did}. Instead, the PNRD relies on a human-in-the-loop to reassess all positive matches to identify false positives (\cite{CJEUCase}, Art. 3 (5) PNRD). The CJEU presumes human verification to be capable of mitigating discriminatory results, without accounting for the base rate fallacy and a high number of false positives that have no discriminatory origin or impact. Consequently, the CJEU's measures to ensure algorithmic accountability and fairness of the PNR system are based on legal fiction.

While the AI Act imposes compliance standards for the providers and deployers of high-risk AI systems, including norms for increasing transparency and countering discriminatory effects, its conditions for accuracy are vague (Arts. 13 (b)(ii); 15 (3), Annex IV (3)) and conditions for robustness are unspecific (Art. 15) \cite{Nolte2024}. The act also perpetuates the secrecy of law enforcement AI (Arts. 43(2), 78 1 (c) (3), 78 (3),  49 (3), 60 (c), Annex  VIII). In sum, although the European rights-based approach recognises the PNR system's impact on fundamental rights, it deems it lawful and legitimate on security grounds, leaving the base rate fallacy unaddressed.  

This article relies on contextual integrity -- a theory on privacy by Helen Nissenbaum \cite{10.5555/1822585} -- as a counterpoint to the rights-based justification of algorithmic surveillance. Instead of relying on fundamental rights to legitimise technology, it allows for identification of relevant normative values bottom-up from the relevant socio-technical context \cite{10.5555/1822585, sanfilippo2018privacy, zimmer2018addressing, sanfilippo2023slow, wernick2023future}. Unlike the European rights-based approach to the protection of privacy and personal data and omnibus legislation, contextual integrity is rooted in the US approach of adopting sector-specific privacy laws \cite{benthall2017contextual, benthallregulatory}, such as HIPAA for healthcare \cite{barth2006privacy}. Contextual integrity can also be deployed  to measure regulations' alignment with the privacy expectations of those affected by them \cite {apthorpe2019evaluating}. This study is the first to quantify passengers' expectations of privacy with respect to the processing of PNR data in the EU, particularly in relation to false positives.

\subsection{Measuring Legitimacy} 
 We conducted a cross-sectional survey of a demographically stratified sample of the adult population of Finland (N=1550) to assess attitudes towards the automated processing of PNR data in air travel and its potential expansion into other travel contexts (for which we used sea travel as an exemplar). Furthermore, focusing on the problem of 
 false positives exposing innocent passengers to rights violations, we developed a novel estimation method that combines a classical vignette study with a preference elicitation method to estimate the false positive rate deemed justifiable by the general public to ensure safety during travel. Our analysis reveals that the perceived appropriateness of algorithmic surveillance is dependent on the travel context, with surveillance in air travel being more acceptable than sea travel. Furthermore, there is a statistically significant difference in what the public views as an acceptable false positive rate between travel contexts, despite the repercussions being the same. Our results also show that the Finnish population is highly tolerant of false positives in general, with legitimate false positive rates of 9.7\% and 14.7\% for sea and air travel, respectively.

Our study shows the feasibility of measuring the contextual integrity of lawful, opaque algorithmic travel surveillance and the value of the contextual integrity perspective as a complementary, albeit not exhaustive, test for legitimacy in rights-based regulatory approaches to privacy, data protection and AI.

\section{Background}

\subsection{Contextual Integrity}

In this empirical study, we rely on Helen Nissenbaum's framework of Contextual Integrity (CI) to describe contexts where PNR technology is introduced. CI is a theory of privacy, whereby it is defined as “a right to \textit{appropriate} flow of information” \cite{10.5555/1822585}. The appropriateness of the informational flow is evaluated in relation to the relevant social context. The CI framework seeks to describe the context-specific informational norms that represent “entrenched expectations governing flows of personal information”. The informational norms reflect the perceived appropriateness of the information flows and are expressed through four parameters: the context, the actors, information types and transmission principles \cite{10.5555/1822585}. 

 At first glance, legal norms may appear to be of limited relevance for context description. For example, empirical studies on CI focus on less regulated environments than the EU and on technologies that data subjects adopt of their own volition, such as smart home systems \cite{abdi2021privacy, apthorpe2018discovering}. However, transmission principles may also be shaped by legislation \cite{barth2006privacy, sanfilippo2020disaster} or companies' privacy policies \cite{shvartzshnaider2019going}. More rarely, the CI framework is used to assess and theorise on the legitimacy of information flows in a regulated environment \cite{selbst2013contextual, apthorpe2019evaluating, guinchard2018taking, hildebrandt2014location, herrmann2016privacy}. Since we deploy the CI lens to assess the legitimacy of a legally mandated surveillance system, legal norms are essential to the description of CI parameters. For example, the PNRD's scope of application defines the relevant context within the broad framework of air travel surveillance. This also allows us to overcome the challenge of context definition associated with applying the CI theory \cite{birnhack2011quest}.

The CI framework can be deployed both to describe and evaluate the context of sharing information and data \cite{10.5555/1822585}. The evaluative use of the CI heuristic has been criticised for its technoconservatism \cite{rule2019contextual, van2022socially} and lack of guidance as to which and whose values matter and how they should be weighted against each other \cite{birnhack2011quest, rule2019contextual, pregent2025you}. Considering that the context studied is normatively framed by the EU law, we have chosen to use the CI framework descriptively as a complement to the rights-based European system of legitimising the use of an algorithmic surveillance system.

The CI framework was developed to describe and assess the contextual integrity of information flows \cite{10.5555/1822585}. As a result, it has been found to be, at times, challenging to apply to incomplete, networked, or aggregate flows of data \cite{benthall2024integrating}, dynamic environments \cite{birnhack2011quest}, opaque contexts and information algorithmically inferred from data \cite{nissenbaum2019contextual, benthallregulatory}. However, the concept of CI and the CI framework are not static \cite{sanfilippo2018privacy, o2022contextual, o2023balancing}, and have been adapted to account for emergent algorithmic systems \cite{nissenbaum2019contextual, benthallregulatory, sileno2021like, oomen2023rethinking, benthall2024integrating}. Although, Prégent has warned against invalid comparisons between two distinct contexts and deemed CI unfit for evaluating disruptive socio-technical contexts \cite{pregent2025you}, in our study, we do not research a disruptive phenomenon but an internationally established travel surveillance practice. We compare the use of the same technology in two different travel environments: air and sea travel. The example of sea travel also references the extension of the PNRD's rules to sea travel in Belgium \cite{CJEUCase}.

Information flows are not always transparent to the key actors. As described in more detail below, some aspects of the PNR system are classified. In other words, it is impossible to describe the technical ``transmission properties'' of these algorithmic systems \cite{benthall2024integrating}. For this reason, we deploy an additional attribute of ``Incomplete Flows'' \cite{shvartzshnaider2019going} to describe what can be known about the algorithmic travel surveillance system. In addition, we seek to specify which actors the system is transparent to \cite{frik2023model}.


\subsection{Context}
This study focuses on the PNR system as it is legally mandated by the PNRD and implemented in Finland, an EU member state, by the national Act 657/2019. The aim of the PNRD is to align EU member states' norms for transmitting and processing PNR data for law enforcement in serious crime and prevention of terrorism. The  EU law and Finnish national legislation determine the information flows relevant to the PNR system.

The PNRD requires analysis of PNR data of all passengers of extra-EU flights (Art. 1). Within the EU, unchecked mass surveillance is not permitted. The PNR analysis can only target selected intra-EU flights when the surveillance is deemed strictly necessary for law enforcement in serious crimes and terrorism (Arts. 1(2); 2(3); \cite{CJEUCase}). Surveillance of all intra-EU flights is permitted only in the presence of a genuine and present or foreseeable terrorist threat \cite{CJEUCase}. In the absence of such a threat, the PNR system can be applied to flights and/or transport operations relating to, \textit{inter alia}, certain travel routes, patterns or nodes for which there are indications justifying application \cite{CJEUCase}. When justified, member states may also apply the PNR analysis to other forms of transport, including sea travel.

Finns' perceptions of the legitimacy of the PNR system may be shaped by their display of high-level trust in the police and other authorities, and expectations of legitimate government use of personal data \cite{OECDFinland}. Finland's corruption perception rate is the second lowest in the world \cite{transparency2023Corruption}.

\subsection{Actors}
The \textit{information subject} is a passenger on an extra-EU flight or a flight or other mode of transport within the EU (PNRD Arts. 1-2, 3(4); \cite{10.5555/1822585}). Air carriers represent \textit{senders of information} (PNRD, Art. 8, \cite{10.5555/1822585}). Passenger information units (PIUs) are the authorities responsible for PNR data processing. They \textit{receive} the PNR data from air carriers or PIUs of other member states and \textit{send} the data or data processing results to competent authorities, such as the police, the customs, or the border guard (PNRD, Art. 6(6), Art. 7), and, upon request, to Europol (PNRD, Art. 10).

\subsection{Attributes}
PNR data encompasses 19 categories of personal data, such as passenger names, all forms of payment information as well as advance passenger information data, including the identity document and nationality (PNRD, Annex 1, 18) \cite{Dtravel}. The inclusion of sensitive categories of data is prohibited (PNRD, Art. 13 (4)). This prohibition aims at limiting the risk of discrimination, although its effectiveness is questioned \cite{korff2021opinion}.

The PNRD also contains rules on the \textit{results of the automated processing} of PNR data, often referred to as the initial ``hit'' \cite{CJEUCase}, and the results of the individual, non-automated review to verify that data (\textit{verified result}). A verified result may be the data on the persons identified or the result of processing that data (PNRD, Art. 6(6)). Although the PNRD is not explicit about this, the PIU also produces information about \textit{the need for further inspection} of the relevant persons.

\subsection{Transmission Principles}
\subsubsection{PNR Data}
The PNRD contains detailed rules on the content, format, and timing of PNR data sharing. Importantly, air carriers must deliver the PNR data to the PIU’s database on the state of departure, landing, or, under certain conditions, the state of the stop-over (Art. 8). 

The PIU processes PNR data to assess passengers before their scheduled arrival or departure from the Member State. The purpose of the processing is to identify persons who require further examination by the competent authorities and, where relevant, by Europol, because such persons may be involved in a terrorist offence or a serious crime (Art. 6).

The PIU must transfer the PNR data of persons identified or the result of processing the PNR data of any positive match to competent authorities in order for them to investigate the involvement of the persons in question in terrorism or serious crime. When relying on automated processing of PNR data, the PIU shall only transmit a verified result.

\subsubsection{Assessment Criteria}
The PIU conducts the assessment automatically by comparing PNR data against databases relevant for preventing, detecting, investigating, and prosecuting terrorist offences and serious crimes (PNRD, Art. 6(3)(a)). According to the Finnish national legislation, names can be compared, for example, against police, border guard and the customs databases and the Schengen Information System. 

Alternatively, the PIU can process the PNR data against pre-determined criteria (PNRD, Art. 6(3)(b)). The assessment must be executed in a non-discriminatory manner with criteria that are targeted, proportionate, and specific (Art. 6(4)). The assessment criteria should be defined in a way that minimises the risk of false positives (PNRD, Rec. 7). The PNRD also contains rules updating and creating the assessment criteria (Art. 6(2)(c)).

Initially, the CJEU ruled that PIUs may not rely on self-learning systems which could alter the pre-determined criteria or weights of the assessment without any human intervention or the opaqueness of which would undermine the right to an effective remedy \cite{CJEUCase}. However, explainable self-learning systems that permit human intervention and assessment could comply with the PNRD \cite{thonnes2022directive}. In contrast to the CJEU's decision, the AI Act permits also the use of self-learning systems AI in the context of  law enforcement, provided that its obligations are complied with (Arts. 3 (1), 8, Annex III, Art. 6 (e)). Given that the AI Act does not affect the application of the EU law on processing personal data and is without prejudice to fundamental rights (Recs. 9-10), the CJEU's guardrails against self-learning PNR systems may prevail.

A human must check any positive result from the automated analysis to verify whether the competent authorities need to act on it \cite{CJEUCase}. The assessment criteria should be clear. The result should not be transmitted where the PIU does not have a reasonable suspicion of the person's involvement in terrorist offences or serious crime with respect to the persons identified by means of those automated processing operations or when they have reason to believe that those processing operations lead to a discriminatory result \cite{CJEUCase}. However, the CJEU did not define what qualifies as a ``true positive'' \cite{korff2023did}. Despite the human-in-the-loop approach, both the results of automated processing and the verified results may contain false positives \cite{thonnes2022directive}.

\subsection{Incomplete Flows}
\subsubsection{Transmission of PNR Data}
 As individuals, passengers have a right to be informed about the processing of their personal data by the airlines in accordance with the GDPR Art. 12 and 13, and should be provided with accurate information that is easy to access and understand about the collection of PNR data, their transfer to PIU and their rights as data subjects (PNRD, Rec. 29).
Generally, the above mentioned information is integrated into the airline’s privacy notice, which covers all processing and sharing of personal data undertaken by the airline \cite{finnair2024privacypolicy}. After passengers’ personal data has been transferred to the competent authorities, the provisions of LED apply (PNRD, Arts. 13(3) and 21(2), LED Rec. 11). Although individuals have a right of access also based on LED (Art. 14), this right can be restricted (Art. 15) on other grounds, such as public and national security.

\subsubsection{Assessment Process and Criteria}
The detailed functioning of the software used by the authorities is considered secret in Finland based on national legislation (\cite{OpennessAct}, Sec. 24(1)(5)). This means that although the individuals would know what personal data the law enforcement authorities process, it is not possible to obtain information on how the PNR software actually makes predictions. The AI Act maintains the confidentiality of AI systems used for law enforcement (Arts. 43(2), 78 1 (c) (3), 78 (3);  49 (3); 60 (c), Annex  VIII C).
Furthermore, the EU member states' reporting obligations (PNRD, Art. 20) do not extend to the rates of false positives and negatives 
(PNRD, Art. 19) \cite{orru2022european}. On this basis, the pre-determined assessment criteria against which the PNR data are evaluated are not public.

Individual passengers become aware of surveillance only if they are stopped at the airport. The initial scrutiny by competent authorities following an automated, positive hit is concealed from them. According to the \textit{Ligue des Droits Humains}, a person becoming aware of being subject to further examination has a right to explanation. During the administrative procedure, the authorities must ensure that the person is able to understand how the pre-determined criteria and PIU's programs work, so that it is possible for that person to decide with full knowledge of the relevant facts whether or not to exercise his or her right to the judicial redress guaranteed in Article 13(1) of the PNR Directive, in order to call into question the possibly unlawful or discriminatory nature of said criteria. However, the explanation should be provided without necessarily allowing that person to become aware of the pre-determined assessment criteria and programs applying those criteria \cite{CJEUCase}.

\section{Approach}


\subsection{Challenges of Opacity} 

EU law and the CJEU have affirmed the legality of the PNR system and established safeguards for fundamental rights protection. Nevertheless, failures in human verification of positive matches can still expose passengers to violations of these rights. While information subjects have the right to information on the processing of their personal data, the secrecy of the assessment process makes it difficult for them to contest the results of surveillance practice. Despite the potential consequences, there is no way for passengers to learn about the system's accuracy or processing principles beyond the limited information provided in airline privacy policies. Moreover, even if detailed information about the PNR system was widely available, the public cannot be expected to understand technical terminology, such as ``false positive'', nor be expected to estimate abstract quantities like a ``false positive rate''. In this sense, both the opacity of the surveillance apparatus and the opacity of the technical nature of the topic make it challenging to assess the public's acceptance of automated surveillance. 

\subsection{Limitations of Vignette Studies}

CI-based research typically assesses the acceptability of information flows using vignette studies (e.g., \cite{zhang2022stop}). In these studies, respondents are presented with hypothetical scenarios where the CI parameters (sender, recipient, information subject, attribute, transmission principle) are systematically varied to identify norm violations, which could, in turn, identify potential privacy violations. These vignettes are randomly sampled for each respondent who indicates their acceptance level using, for example, a 5-point Likert response scale. The responses are then pooled for analysis \cite{heise2010surveying}.

In our case, we wanted to assess the distribution of acceptability that different numbers of false positives have in the PNR system. We could similarly create vignettes based on specific numbers of false positives and ask survey respondents to rate their acceptability. However, this approach has several limitations. First, Likert response scales are too course-grained to distinguish between the acceptability of consecutive false positive values, limiting us to identifying a broad range of values the public would deem acceptable. Second, it is sample inefficient: if a respondent states, for example, that 10 false positives are acceptable, then it implies something about nearby values that we would ideally like to incorporate into our analysis. Last, and most importantly, we specifically care about the trade-off that the number of false positives represents, which is difficult to highlight in a conventional vignette study.

\subsection{Critiquing Outcomes with Collective Criticism}

To overcome these challenges, we created a novel vignette study using the collective criticism approach \cite{medlar2022critiquing, medlar2017} to apply the CI framework to a context qualified by multilayered opacity. Collective criticism was originally designed to model trade-offs based on user preferences in AI and interactive systems. Here we created a vignette related to false positives in the PNR system and used the feedback and analysis methodologies from collective criticism to estimate the threshold, beyond which, it breaches passengers' perceptions of contextual integrity. 

In brief, respondents were shown a ``typical'' example of passenger pre-screening resulting in a given (randomly sampled) number of false positives. Respondents were then asked to make a trade-off between security and privacy. If respondents prioritised security, then a higher number of false positives would be acceptable, whereas prioritising privacy means that the number of false positives should be lower. These responses are combined with the number of false positives shown to the respondent and transformed into intervals. These intervals are then analysed using interval regression along with other predictor variables, such as the travel context. This procedure is described in more detail in the next section.

\section{Methodology}

We conducted a nationally representative survey study to address the following research question: ``\textit{How do Finnish citizens perceive the legitimacy of algorithmic surveillance in EU cross-border travel?}'' In particular, we sought to 
\begin{enumerate*}[label=(\roman*)]
    \item estimate what Finns perceived as a justifiable false positive rate when identifying criminals and terrorists at the border, and 
    \item provide supportive evidence for this estimate by drawing from their attitudes towards the use of personal data in automated passenger pre-screening.
\end{enumerate*}

\subsection{Cross-Sectional Survey}

We designed a cross-sectional survey where respondents were randomly assigned to one of two experimental conditions: air travel or sea travel. Both experimental conditions were identical other than changes in terminology (e.g., airport versus ferry terminal).
The survey contained demographic questions (8 items), attitudinal questions related to the use of passenger data and automated computer analysis in cross-border travel (12 items), and a vignette where respondents were asked to make a trade-off between security and privacy in passenger pre-screening based on a given number of false positives (1 item).
The survey was administered in Finnish. The working languages when creating the questionnaire were Finnish and English, with the final Finnish version being checked by a language specialist. For clarity, in this article we present the survey items in idiomatic English. However, both English and Finnish versions of the survey items are provided in the appendix.

\subsubsection{Demographic Questions}

We collected demographic and other background information to ensure that each experimental condition was comparable. The questions were related to respondents' gender, age group, level of education, whether they self-identified as a member of an ethnic group discriminated against in Finland, typical frequency and purpose of travel (by air or sea, dependent on experimental condition), and whether they had been afraid of becoming a victim of crime or involved in a terrorist attack within the last year (see Supplementary Table~\ref{tab:demographics_dual}).

\subsubsection{Attitudes towards Automated Passenger Pre-screening}

We investigated three aspects of respondents' attitudes to automated passenger pre-screening: 
\begin{enumerate*}[label=(\roman*)]
    \item usage of passenger data,
    \item usage of automated computer analysis, and 
    \item the consequences of automated computer analysis (see Table~\ref{tab:attitudes} and Supplementary Table~\ref{tab:attitudes_dual} for English and Finnish translations, respectively).
\end{enumerate*}
All survey items used a 5-point Likert response scale ({\em strongly disagree -- strongly agree}) together with an ``{\em I don't know}'' option, with the exception of Q10 where we used binary responses to assess the acceptability of various privacy violations that could result from individuals being flagged  during pre-screening.

As the survey was created with the goal of being administered to a nationally representative sample of the adult population of Finland, we aimed to use simple descriptive terms that assumed no technical knowledge and provided examples where necessary. For example, we refer to specific law enforcement agencies by name, rather than use blanket terms like ``the authorities'' (see, for example, Q1) and provided examples of travel patterns thought to resemble those of criminals (see Q8).

\begin{table*}[t]
    \centering
    \footnotesize
    \caption{Attitudinal questions on automated passenger pre-screening. Square brackets indicate alternate wordings for air and sea travel experimental conditions.}
    \renewcommand{\arraystretch}{1.1}
    \begin{tabular}{l p{0.95\textwidth}} \toprule
        \multicolumn{2}{c}{Usage of passenger data} \\ \midrule
        Q1 & It is acceptable for the [airline/ferry company] to provide the police, customs and border guard with passenger data for the purpose of crime prevention.\\
        Q2 & It is acceptable for the [airline/ferry company] to provide Valvira (National supervisory authority for welfare and health) with passenger data for the purpose of preventing infectious disease transmission during a pandemic. \\
        Q3 & It is acceptable for the police, customs, and the border guard to use passenger data to automatically pre-screen passengers for potential terrorists and criminals. \\
        Q4 & In addition to passenger data, I would willingly give additional personal information (e.g., social media posts or credit card purchase information) to the police, customs, and the border guard, if it would make security checks faster and more convenient.\\
        Q5 & The use of my passenger data by the police, customs and the border guard makes me feel that the state does not trust me.\\ \midrule
        \multicolumn{2}{c}{Usage of automated computer analysis} \\ \midrule
        Q6 & It is acceptable to use passenger data for automated pre-screening of all passengers to identify potential criminal threats. \\
        Q7 & I would rather my passenger data be pre-screened automatically by a computer than manually by a human.\\
        Q8 & It is acceptable to flag passengers as suspicious if they exhibit travel patterns that resemble those of criminals (e.g., last minute bookings, payment in cash, unusually long or expensive routes, no baggage). \\
        Q9 & Automated pre-screening of all passengers in [air/sea] travel based on passenger data is an excessive use of state power. \\ \midrule
        \multicolumn{2}{c}{Consequences of automated computer analysis} \\ \midrule
        Q10 & It is acceptable for police officers to manually examine the following information about individuals who have been flagged during passengers pre-screening: \newline
        \hphantom{1} a) Information on police databases (e.g., connections to ongoing criminal investigations) \newline
        \hphantom{1} b) Criminal record \newline
        \hphantom{1} c) Information in the Population Information System (including personal and family data, details of property and building ownership, and previous name(s) and address(es), \newline
        \hphantom{1} current and previous memberships of a religious community) \newline
        \hphantom{1} d) MyKanta (containing health records and prescriptions) \newline
        \hphantom{1} e) Credit register (containing information on defaulted payments) \newline
        \hphantom{1} f) Social media \newline
        \hphantom{1} g) News about the person that can be found online \newline
        \hphantom{1} h) None of the above are acceptable 
        \\
        Q11 & It is acceptable for police officers to stop flagged passengers at the [airport/ferry terminal] to confirm whether the person is an actual criminal threat. \\
        Q12 & It is acceptable for police officers to stop flagged passengers at the [airport/ferry terminal] to confirm whether the person is an actual criminal threat even if it causes them to miss their [flight/departure].\\ \bottomrule
    \end{tabular}
    \label{tab:attitudes}
\end{table*}



\subsubsection{False Positives from Pre-screening Vignette}

We used collective criticism -- a vignette-based approach for modelling trade-offs -- to estimate the legitimate false positive rate during automated passenger pre-screening \cite{medlar2022critiquing}.
In the vignette, we state that a subset of 200 individuals were automatically flagged during passenger pre-screening. The personal information of flagged passengers was then manually examined by law enforcement officers to reclassify them into true positives (e.g., criminals with outstanding arrest warrants) and false positives (innocent passengers). We referred to false positives as \textit{false alarms}. 
Respondents were then asked to reflect on the scenario and answer the following question: 

{\em Given the number of false alarms presented above, which statement are you more in agreement with:}
\begin{enumerate}
    \item \textit{I prioritise identifying criminal threats over the privacy of innocent passengers, even if it means more false alarms and privacy violations of innocent passengers by the police.}
    \item \textit{I prioritise protecting innocent passengers’ privacy, even if reducing false alarms also means fewer alerts of criminal threats.}
\end{enumerate}

\begin{figure}
    \centering
    \includegraphics[width=0.5\textwidth]{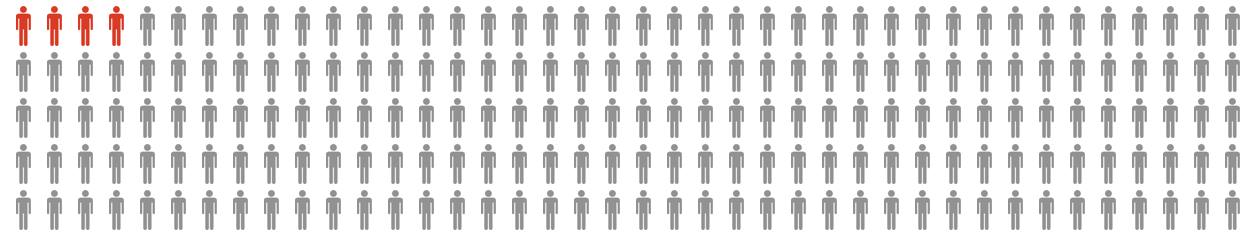}
    \caption{Example of image used in the false positives vignette question. 
    Respondents were told that red passengers had been automatically flagged during passenger pre-screening.}
    \Description{A grid of 200 stick figures representing passengers. Ten stick figures are coloured red to indicate passengers flagged during pre-screening, the remainder are coloured grey.}
    \label{fig:stickmen}
\end{figure}

\begin{figure}
    \centering
    \includegraphics[width=0.46\textwidth]{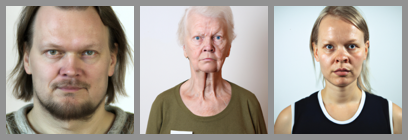}
    \caption{Example passport photos used in the false positives vignette question.}
    \Description{Four passport photo images generated by DALL-E 2. Each image shows adults of different ages and genders looking into the camera.}
    \label{fig:faces}
\end{figure}

We generated 150 vignettes per experimental condition by varying the number of true positives and false positives presented to the respondents (true positives: 0-4, false positives: 1-30). These ranges were selected because true positives are assumed to be rare, but the number of false positives viewed as legitimate by the general public is unknown, so a wider range was chosen. 
We included a visualisation of the proportion of flagged passengers (see Figure~\ref{fig:stickmen}) and generated synthetic passport photos using DALL-E 2 \cite{ramesh2022hierarchical} to humanise the passengers classified as true and false positives (see Figure~\ref{fig:faces}). 
The images were shown inline at relevant points in the vignette.

The vignette was split over 4 pages to carefully walk respondents through the details of the scenario and to better accommodate mobile devices. The complete scenario is provided in the appendix (see Supplementary Table~\ref{tab:vignette_dual}).

\subsection{Survey Deployment}

The survey was conducted by Taloustutkimus\footnote{\url{https://www.taloustutkimus.fi/}}, a Finnish market research company, using their Internet research panel from 12-28th April 2023. Taloustutkimus' Internet panel was recruited using offline (e.g., telephone, mail) and online (e.g., emails, social media) methods. The surveys for air travel and sea travel were distributed to a nationally representative sample of the adult population of Finland using stratified sampling on the basis of gender (female, male), age (13 age brackets) and region (Northern and Eastern Finland, Western Finland, Southern Finland and Helsinki). While we recognise the importance of including non-binary genders, this information was unavailable during sampling. 

For the air travel survey, 5058 panellists were invited to participate and 772 (15.3\%) completed the survey. For the sea travel survey, 6125 panellists were invited to participate and 778 (12.7\%) completed the survey. According to Taloustutkimus, these response rates are typical for nationally representative surveys in Finland. The mean time respondents spent filling out the survey was $\mathord{\sim}$7.5 minutes. Based on the number of respondents, the margin of error was 3.5\% at a 95\% confidence interval for each experimental condition.

\subsection{Data Analysis}

In our survey, we measured attitudes towards automated passenger pre-screening using 5-point Likert and binary response scales. 

\subsubsection{Sample Balancing}

We used sample balancing to reduce biases resulting from non-response, over-representation and other coverage issues. Sample balancing was performed using population estimates provided by Statistics Finland from 2022 according to age and gender by region. Post-processing and data analysis were performed using the Survey R package (version 4.4.2) \cite{lumley2004analysis}. 

\subsubsection{Attitudinal Survey Questions}

We analysed each survey item as a distinct outcome to identify differences in respondents' attitudes to automated passenger pre-screening between experimental conditions. As survey responses are not normally distributed, we used non-parametric statistical tests, including $\chi^2$ and Wilcoxon rank-sum tests for survey items with binary and Likert response scales, respectively\footnote{We note that rank-sum tests are implemented using $t$-statistics for sample balanced survey data and, therefore, are reported in the same manner as $t$-tests \cite{lumley2013two}}. 

\subsubsection{False Positive Rate Estimation}

Following Medlar et al.~\cite{medlar2022critiquing}, we modelled the acceptable number of false positives by transforming responses 
into intervals. For example, if a respondent was shown a scenario with 10 false positives (out of 200 passengers) and they prioritised the identification of criminal threats, then the optimal number of false positives would lie in the interval (10, 200]. On the other hand, if they had prioritised privacy, then we would transform their response into the interval [0, 10).

We fitted interval regression models using the intervals derived above as the dependent variable. As the number of false positives cannot be negative, we assumed the dependent variable had a log-normal distribution\footnote{Equivalent to a log transform of the dependent variable in linear regression.}. We additionally considered the experimental condition (air or sea travel), respondents' gender, age group, and the number of true positives as potential independent variables during model selection. As with the attitudinal questions, the data was based on a nationally representative sample, so sample balancing was performed prior to fitting regression models.

\subsection{Limitations}

Our study has several limitations. 
First, we do not investigate false negatives -- where a criminal or terrorist is not flagged by the surveillance system -- because the determination of false negatives is only possible in a controlled evaluation where we have access to the ground truth and not in a deployed system. We acknowledge, however, that were false negatives known to occur (e.g., after a successful terrorist attack), then respondents may be more likely to question the legitimacy of the current procedures for passenger pre-screening.
Second, we assume that the process of human verification of positive matches -- the human-in-the-loop -- never fails to reclassify an innocent passenger as a false positive. We made this simplifying assumption as there is no available information about the accuracy of verification, but we are aware that this represents the base-case scenario for correcting errors in automated surveillance.


\section{Results}

To address our research question, we first compare the sample characteristics of each experimental condition to identify potential biases. Next, we show that Finns are generally tolerant of false positives in automated surveillance, but that perceived legitimacy of false positives is dependent on both the travel context and gender. Last, we support these estimates by drawing on respondents' attitudes to various aspects of automated passenger pre-screening.

\subsection{Sample Characteristics}


Participants were representative of the adult population of Finland in terms of 
gender (Female: unweighted percentage 49.9\% vs benchmark 50.8\%) and 
education (Tertiary education: unweighted percentage 30\% vs benchmark 33.3\%).
Self-described membership of an ethnic group discriminated against in Finland was low (2.1\%).
The typical number of trips taken in a (non-pandemic) year was also low (M=1.51, SD=2.71), with the most common purpose being private reasons (i.e., unrelated to work). A majority of respondents reported not being afraid of becoming a victim of a crime (unweighted percentage, 76.6\%) or a terrorist attack (unweighted percentage, 93.1\%) in the last year. 

We compared background questions (excluding those used for sample balancing) to verify similarity across experimental conditions. There were no significant differences in the level of education, membership of an ethnic group discriminated against in Finland, typical frequency of travel, and fear of becoming the victim of a crime or a terrorist attack for respondents answering the air travel and sea travel surveys ($p > 0.05$). There was, however, a significant difference in the main purpose of travel, with air travel being more common for work than sea travel ($\chi^2(2) = 15.77, p < 0.001$). 

\subsection{Perceived Legitimacy of False Positives is Influenced by Travel Context and Gender}

We investigated whether four factors influenced respondents' views of false positives in algorithmic surveillance: age, gender, travel context and the number of true positives shown in the vignette.
Each factor was assessed using ANOVA to compare nested regression models (i.e., two regression models where a given factor is missing from one of the models).

We found no evidence that  
the age of respondents $(F(1,1545)=0.40, p=0.529)$ or 
the number of true positives $(F(1,1545)=0.65, p=0.420)$
affected respondents' acceptance of false positives.
However, whether respondents were randomly assigned to air travel or sea travel ($F(1,1545)=4.56, p=0.033$) and their gender $(F(1,1545)=10.85, p=0.001)$ had a significant impact on responses. 
We, therefore, modelled the legitimate number of false positives on the basis of travel context and gender.


\subsection{Higher Perceived Legitimacy of False Positives in Air Travel than Sea Travel}

\begin{table}[t]
\centering
\caption{Estimates for the legitimate number of false positives that balance security and the privacy of innocent passengers based on mode of travel and respondents' gender ($\ast$ = false positive rate calculation based on a total of 200 passengers).}
\resizebox{0.48\textwidth}{!}{%
\begin{tabular}{llccl} \toprule
\multirow{2}{*}{\begin{tabular}[c]{@{}l@{}}Travel\\ mode\end{tabular}} &
  \multirow{2}{*}{Gender(s)} &
  \multirow{2}{*}{\begin{tabular}[c]{@{}c@{}}Legitimate\\ False Positives\end{tabular}} &
  \multirow{2}{*}{\begin{tabular}[c]{@{}c@{}}Confidence\\ interval (95\%)\end{tabular}} &
  \multirow{2}{*}{FPR*} \\
                     &        &       &                    &       \\ \midrule
\multirow{3}{*}{Air} & Male+female      & 25.83 & {[}23.18, 28.47{]} & 0.129 \\
                     & Male   & 22.85 & {[}19.88, 25.82{]} & 0.114 \\
                     & Female & 29.42 & {[}25.80, 33.03{]} & 0.147 \\ \midrule
\multirow{3}{*}{Sea} & Male+female    & 22.10 & {[}19.58, 24.63{]} & 0.111 \\
                     & Male   & 19.37 & {[}16.55, 22.19{]} & 0.097 \\
                     & Female & 24.94 & {[}21.77, 28.11{]} & 0.125 \\ \bottomrule
\end{tabular}%
}
\label{tab:fpr}
\end{table}

Table~\ref{tab:fpr} shows the estimates for the legitimate number of false positives, in addition to their corresponding 95\% confidence intervals and false positive rates based on a total of 200 passengers (as was shown to respondents in Figure~\ref{fig:stickmen}).
Respondents made different privacy trade-offs based on the travel context and their gender. 
Respondents placed more emphasis on identifying criminal threats for air travel than sea travel (25.83 versus 22.10 false positives, respectively).
We found a similar pattern across female and male respondents. After controlling for gender, the optimal number of false positives was 28.8\% higher for women than for men for both air and sea travel.
We note, however, that despite the differences between trade-offs made by the respondents, even the lowest optimal number of false positives was still high (19.37 false positives for male respondents and sea travel), corresponding to a false positive rate of almost 10\%.
We present the full table of regression coefficients in the Appendix (see Supplementary Tables~\ref{tab:regression_travel_gender} and \ref{tab:regression_travel_only}).

\subsection{Attitudes toward Automated Pre-screening are Consistent with Perceived Legitimacy of False Positives}

Respondents' perceived legitimacy of false positives was broadly consistent with their attitudes towards other aspects of automated passenger pre-screening, where they were more likely to find the use of passenger data, and the use and consequences of automated computer analysis more acceptable in air travel than sea travel.

In this section, we do not control for gender as randomisation ensures that variables, such as gender and age group, are not confounders. In the previous analyses, we controlled for gender because it was found to have a moderating effect on the perceived legitimacy of false positives.

\subsubsection{Usage of Passenger Data}

Respondents found it more acceptable for the police, customs, and border guard to use passenger data for automatic passenger pre-screening for potential terrorists and criminals for air travel than for sea travel (Q3, $t(1525) = -2.56, p = 0.011$), 
and were more willing to give the authorities additional personal information to speed up the process (Q4, $t(1473) = -2.70, p = 0.007$). 
Moreover, respondents were less likely to interpret the use of passenger data as a sign of mistrust on the part of the state during air travel than sea travel (Q5, $t(1515) = 2.42, p = 0.016$).

\subsubsection{Usage of Automated Computer Analysis}

Respondents found the use of automated computer analysis to be more acceptable in air travel than sea travel in every aspect that we investigated.
In air travel, it is more acceptable to use automated computer analysis to identify potential criminal threats (Q6, $t(1502) = -2.81, p = 0.005$) and to flag passengers on the basis of travel patterns (Q8, $t(1481) = -3.27, p = 0.001$). 
Whereas, in sea travel, respondents found it less acceptable for pre-screening to be conducted by humans (Q7, $t(1399) = 2.37, p = 0.037$) and 
were more likely to agree that automated pre-screening was an excessive use of state power (Q9, $t(1472) = 2.83, p = 0.005$).

\subsubsection{Consequences of Automated Computer Analysis}

In general, respondents found it more acceptable for the police to manually examine the personal information of passengers flagged during pre-screening during air travel than sea travel. This included manual examination of 
the population information system (Q10c, $\chi^2(1) = 20.83, p < 0.001$),
health records (Q10d, $\chi^2(1) = 11.87, p = 0.003$),
the credit register (Q10e, $\chi^2(1) = 12.33, p = 0.001$),
and online searches about the individuals in question (Q10g, $\chi^2(1) = 10.28, p = 0.002$).
It was also more acceptable for the police to stop flagged passengers during air travel than during sea travel to confirm whether they posed a genuine criminal threat (Q11, $t(1517) = -2.20, p = 0.028$). However, if this confirmation caused passengers to miss their departures, 
then there was no difference in acceptability between the travel contexts (Q12, $p > 0.05$).

\section{Discussion}
Due to the base rate fallacy, even extremely accurate algorithmic mass surveillance systems are guaranteed to produce a 
high number of false positives, exposing their targets to violations of rights to privacy, data protection, non-discrimination and effective judicial remedy (Arts. 7, 8, 21, 47 CFREU). 
Indeed, the EU judiciary and legislator struggle to account for the base rate fallacy associated with algorithmic mass surveillance \cite{korff2023did, korff2021opinion, thonnes2022directive, Thönnes}, despite the EU's reputation for a strict rights-based approach to technology regulation \cite{bradford2023digital}. The CJEU permits opaque algorithmic PNR surveillance, relying on humans-in-the-loop and conditions on the deployment of self-learning systems to control for false positives \cite{CJEUCase}. The CJEU's failure to address problems related to the base rate fallacy in \textit{Ligue des Droits Humains} leaves a gap in the legitimacy of the PNR system, illustrating how top-down, rights-based approaches are vulnerable to lawmakers and judges either failing to understand the technical and statistical properties of a surveillance system or deliberately refraining from stating the reasons for legitimising the use of a system producing a high number of false positives. 

We deploy the theory of contextual integrity \cite{10.5555/1822585} to complement the legitimacy gap observed in the rights-based approaches in legitimising PNR surveillance. We focus on the perception of a privacy-security trade-off, a matter which had not previously been studied by national and EU authorities. However, it is challenging to study the public perception of automated surveillance due to its physical and legal opacity as well as challenges in communicating the meaning of ``false positives'' and ``false negatives.'' In response, we deployed a novel vignette-based estimation method to assess passengers' acceptance of such a surveillance system and to assess where the false positive rates breached the perceived threshold of being legitimate. 

Our respondents displayed a consistently higher acceptance rate of surveillance in air compared to sea travel. Air travel is routinely surveilled, and passengers can be expected to have become accustomed to relinquishing some of their privacy in favour of safety and security \cite{10.5555/1822585}. The lower acceptance rate of sea travel surveillance may reflect resistance to ``function creep'' from air to sea travel. The novelty in our experiment was the 
estimation of the distribution of acceptability 
of false positives in algorithmic mass surveillance, which was surprisingly high in both travel contexts. 
This high rate may be explained by the Finnish population's high trust in law enforcement authorities and their processing of personal data \cite{OECDFinland}, whereas the difference between genders could be related to the high degree of gender-based violence in Finland\footnote{\url{https://eige.europa.eu/gender-equality-index/2023/domain/violence/FI}}. Of course, the threshold for breaching the contextual integrity of the privacy-security trade-off may be different in contexts with lower trust in police and public authorities, or higher degrees of ethnic diversity.
We acknowledge that respondents' preferences with respect to false positives could also be confounded by normalcy bias, where individuals minimise the personal risks associated with, in this case, privacy violations that have the potential to impact their lives (i.e., ``this is something that happens to other people''). This could confound the magnitude of the estimated false positive rate, but not the differences between travel contexts. While this raises questions about people’s capacity to account for privacy harms that happen to other people, it is a limitation that affects both rights-based and contextual integrity-based approaches.
From a practical perspective, it is unclear whether the existing PNR system could scale to accommodate false positive rates in excess of 10\%, especially given its reliance on the manual inspection of individual passengers. Indeed, based on our estimate of the justifiable false positive rate in air travel of 12.7\%, there would need to be an aggregate capacity of $\mathord{\sim}140,000$ manual checks per day at the EU-level\footnote{Based on Eurostat figures of 198M total air passengers carried at the EU-level from January-March 2024, of which 50.1\% were on extra-EU flights, \url{https://ec.europa.eu/eurostat/statistics-explained/index.php?title=Air_passenger_transport_-_monthly_statistics&oldid=616429}.}.


 In light of the limitations of the CI approach, it is essential to reflect on how rights-based approaches account for the base rate fallacy as a statistical property of algorithmic mass surveillance. The issue is non-trivial, considering other proposals for mass surveillance, such as the EU Commission's proposal for monitoring online communications to combat child sexual abuse material \cite{Thönnes, Commissionprposal}. The justifications for deploying algorithmic surveillance should be based on the right premises. To an extent, it is a matter of effective communication. The term ``false positive'' is ambiguous, leading to misunderstandings between different stakeholders \cite{darling2021you}. The risks related to the base rate fallacy should be better understood in the legal community and better communicated between producers and deployers of algorithmic surveillance systems. The general public should be informed about the systems' false positive rates. The AI Act's ambiguous norms for accuracy and robustness \cite{Nolte2024} should be evaluated for their capacity to account for the base rate fallacy. One can interpret the Act to require the producers of the AI systems to inform system deployers about the high numbers of false positives stemming from the base rate fallacy (AI Act, Art. 9(5) and 13(3)(ii)).

To conclude, the CI framework can also be relied upon to complement rights-bases approaches to legitimise algorithmic surveillance. However, neither the top-down nor the bottom-up approach is without weaknesses, with the former demanding lawmakers and judges to understand the technology and the latter requiring it from the general public, who may display normalcy bias. It must be acknowledged that the CI theory, and our study, may not account for all power relations present in the relevant context \cite{birnhack2011quest}. However, this limitation also extends to rights-based approaches. Finally, both the CI and rights-based approaches focus on the individual \cite{Yeungcoe, smuha2021beyond, benthall2024integrating, benthallregulatory} and, as a result, may not fully account for the systemic and collective effects of mass surveillance. Here, policy-making and socio-legal research should remain informed by surveillance studies research.




\section*{Positionality Statement}

Three out of four authors are first-generation immigrants in Finland. Our interpretation of the quantitative results in this work may, therefore, have been influenced by our minority status within Finnish society.

\begin{acks}
This work has been supported by the Helsinki Institute for Information Technology HIIT through a community support grant, the Kone Foundation, Finland through the ``Long-term human rights risks of smart city technologies'' project at the Legal Tech Lab, The Faculty of Law, University of Helsinki, and the Carl Zeiss Foundation. Alina Wernick is a member of the Machine Learning Cluster of Excellence, EXC number 2064/1 – Project number 390727645. 

We would like to thank Riikka Koulu for feedback on the different stages of the research design and earlier manuscripts; Kati Nieminen, Suvi Sankari and Veera Koponen for their feedback on the design of the questionnaire and the empirical study;  Kimmo Nuotio and Janne Kivivuori for their feedback at the research seminars on criminal law and criminology; Kristof Meding for giving feedback of the earlier manuscript and highlighting the relevance of the base rate fallacy;  Markus Ahlers on the feedback on the earlier manuscript and integration of the relevant literature streams; Sebastian Benthall and Madelyn R. Sanfilippo for their invaluable feedback and insight on the application of the CI lens;  Michèle Finck for her advice on a suitable interdisciplinary publication outlet; Tharushi Abeynayaka on the background research;  Emeline Banzuzi for advice on the literature on algorithmic discrimination. 
\end{acks}

\balance
\bibliographystyle{ACM-Reference-Format}
\bibliography{sample-base}


\begin{thebibliography}{79}


\ifx \showCODEN    \undefined \def \showCODEN     #1{\unskip}     \fi
\ifx \showDOI      \undefined \def \showDOI       #1{#1}\fi
\ifx \showISBNx    \undefined \def \showISBNx     #1{\unskip}     \fi
\ifx \showISBNxiii \undefined \def \showISBNxiii  #1{\unskip}     \fi
\ifx \showISSN     \undefined \def \showISSN      #1{\unskip}     \fi
\ifx \showLCCN     \undefined \def \showLCCN      #1{\unskip}     \fi
\ifx \shownote     \undefined \def \shownote      #1{#1}          \fi
\ifx \showarticletitle \undefined \def \showarticletitle #1{#1}   \fi
\ifx \showURL      \undefined \def \showURL       {\relax}        \fi
\providecommand\bibfield[2]{#2}
\providecommand\bibinfo[2]{#2}
\providecommand\natexlab[1]{#1}
\providecommand\showeprint[2][]{arXiv:#2}

\bibitem[Abdi et~al\mbox{.}(2021)]%
        {abdi2021privacy}
\bibfield{author}{\bibinfo{person}{Noura Abdi}, \bibinfo{person}{Xiao Zhan}, \bibinfo{person}{Kopo~M Ramokapane}, {and} \bibinfo{person}{Jose Such}.} \bibinfo{year}{2021}\natexlab{}.
\newblock \showarticletitle{Privacy norms for smart home personal assistants}. In \bibinfo{booktitle}{\emph{Proceedings of the 2021 CHI conference on human factors in computing systems}}. \bibinfo{pages}{1--14}.
\newblock


\bibitem[Apthorpe et~al\mbox{.}(2018)]%
        {apthorpe2018discovering}
\bibfield{author}{\bibinfo{person}{Noah Apthorpe}, \bibinfo{person}{Yan Shvartzshnaider}, \bibinfo{person}{Arunesh Mathur}, \bibinfo{person}{Dillon Reisman}, {and} \bibinfo{person}{Nick Feamster}.} \bibinfo{year}{2018}\natexlab{}.
\newblock \showarticletitle{Discovering smart home internet of things privacy norms using contextual integrity}.
\newblock \bibinfo{journal}{\emph{Proceedings of the ACM on interactive, mobile, wearable and ubiquitous technologies}} \bibinfo{volume}{2}, \bibinfo{number}{2} (\bibinfo{year}{2018}), \bibinfo{pages}{1--23}.
\newblock


\bibitem[Apthorpe et~al\mbox{.}(2019)]%
        {apthorpe2019evaluating}
\bibfield{author}{\bibinfo{person}{Noah Apthorpe}, \bibinfo{person}{Sarah Varghese}, {and} \bibinfo{person}{Nick Feamster}.} \bibinfo{year}{2019}\natexlab{}.
\newblock \showarticletitle{Evaluating the Contextual Integrity of Privacy Regulation: Parents' {IoT} Toy Privacy Norms Versus {COPPA}}. In \bibinfo{booktitle}{\emph{28th USENIX security symposium (USENIX security 19)}}. \bibinfo{pages}{123--140}.
\newblock


\bibitem[Bar-Hillel(1980)]%
        {bar1980base}
\bibfield{author}{\bibinfo{person}{Maya Bar-Hillel}.} \bibinfo{year}{1980}\natexlab{}.
\newblock \showarticletitle{The base-rate fallacy in probability judgments}.
\newblock \bibinfo{journal}{\emph{Acta Psychologica}} \bibinfo{volume}{44}, \bibinfo{number}{3} (\bibinfo{year}{1980}), \bibinfo{pages}{211--233}.
\newblock


\bibitem[Barth et~al\mbox{.}(2006)]%
        {barth2006privacy}
\bibfield{author}{\bibinfo{person}{Adam Barth}, \bibinfo{person}{Anupam Datta}, \bibinfo{person}{John~C Mitchell}, {and} \bibinfo{person}{Helen Nissenbaum}.} \bibinfo{year}{2006}\natexlab{}.
\newblock \showarticletitle{Privacy and contextual integrity: Framework and applications}. In \bibinfo{booktitle}{\emph{2006 IEEE symposium on security and privacy}}. IEEE.
\newblock


\bibitem[Benthall and Cummings(2024)]%
        {benthall2024integrating}
\bibfield{author}{\bibinfo{person}{Sebastian Benthall} {and} \bibinfo{person}{Rachel Cummings}.} \bibinfo{year}{2024}\natexlab{}.
\newblock \showarticletitle{Integrating differential privacy and contextual integrity}. In \bibinfo{booktitle}{\emph{Proceedings of the Symposium on Computer Science and Law}}. \bibinfo{pages}{9--15}.
\newblock


\bibitem[Benthall et~al\mbox{.}(2017)]%
        {benthall2017contextual}
\bibfield{author}{\bibinfo{person}{Sebastian Benthall}, \bibinfo{person}{Seda G{\"u}rses}, \bibinfo{person}{Helen Nissenbaum}, {et~al\mbox{.}}} \bibinfo{year}{2017}\natexlab{}.
\newblock \showarticletitle{Contextual integrity through the lens of computer science}.
\newblock \bibinfo{journal}{\emph{Foundations and Trends{\textregistered} in Privacy and Security}} \bibinfo{volume}{2}, \bibinfo{number}{1} (\bibinfo{year}{2017}), \bibinfo{pages}{1--69}.
\newblock


\bibitem[Benthall and Sivan-Sevilla({[n.\,d.]})]%
        {benthallregulatory}
\bibfield{author}{\bibinfo{person}{Sebastian Benthall} {and} \bibinfo{person}{Ido Sivan-Sevilla}.} \bibinfo{year}{[n.\,d.]}\natexlab{}.
\newblock \bibinfo{title}{Regulatory CI: Adaptively Regulating Privacy as Contextual Integrity}.
\newblock
\newblock


\bibitem[Birnhack(2011)]%
        {birnhack2011quest}
\bibfield{author}{\bibinfo{person}{Michael Birnhack}.} \bibinfo{year}{2011}\natexlab{}.
\newblock \showarticletitle{A Quest for a Theory of Privacy: Context and Control’(2011)}.
\newblock \bibinfo{journal}{\emph{Jurimetrics}}  \bibinfo{volume}{51} (\bibinfo{year}{2011}), \bibinfo{pages}{447}.
\newblock


\bibitem[Bradford(2023)]%
        {bradford2023digital}
\bibfield{author}{\bibinfo{person}{Anu Bradford}.} \bibinfo{year}{2023}\natexlab{}.
\newblock \bibinfo{booktitle}{\emph{Digital empires: The global battle to regulate technology}}.
\newblock \bibinfo{publisher}{Oxford University Press}.
\newblock


\bibitem[Brouwer et~al\mbox{.}(2023)]%
        {brouwer2023future}
\bibfield{author}{\bibinfo{person}{Evelien Brouwer}, \bibinfo{person}{Elspeth Guild}, \bibinfo{person}{Stefan Salomon}, {and} \bibinfo{person}{Christian Th{\"o}nnes}.} \bibinfo{year}{2023}\natexlab{}.
\newblock \showarticletitle{The Future of the European Security Architecture-a Debate Series}.
\newblock \bibinfo{journal}{\emph{Originally published on: https://verfassungsblog. de/category/debates/pnr-debate-series/, Max Planck Institute for the Study of Crime, Security and Law Working Paper}} \bibinfo{number}{2023-06} (\bibinfo{year}{2023}).
\newblock


\bibitem[Browning and Arrigo(2021)]%
        {browning2021stop}
\bibfield{author}{\bibinfo{person}{Matthew Browning} {and} \bibinfo{person}{Bruce Arrigo}.} \bibinfo{year}{2021}\natexlab{}.
\newblock \showarticletitle{Stop and risk: Policing, data, and the digital age of discrimination}.
\newblock \bibinfo{journal}{\emph{American Journal of Criminal Justice}} \bibinfo{volume}{46}, \bibinfo{number}{2} (\bibinfo{year}{2021}), \bibinfo{pages}{298--316}.
\newblock


\bibitem[CJEU(2022)]%
        {CJEUCase}
\bibfield{author}{\bibinfo{person}{CJEU}.} \bibinfo{year}{2022}\natexlab{}.
\newblock \bibinfo{title}{Case C-817/19, Ligue des droits humains}.
\newblock
\newblock


\bibitem[Commission(2022)]%
        {Commissionprposal}
\bibfield{author}{\bibinfo{person}{European Commission}.} \bibinfo{year}{2022}\natexlab{}.
\newblock \bibinfo{title}{Proposal for a REGULATION OF THE EUROPEAN PARLIAMENT AND OF THE COUNCIL laying down rules to prevent and combat child sexual abuse}.
\newblock
\newblock
\urldef\tempurl%
\url{https://eur-lex.europa.eu/legal-content/EN/TXT/HTML/?uri=CELEX:52022PC0209}
\showURL{%
\tempurl}


\bibitem[{Council of European Union}(2000)]%
        {CFREU}
\bibfield{author}{\bibinfo{person}{{Council of European Union}}.} \bibinfo{year}{2000}\natexlab{}.
\newblock \bibinfo{title}{Charter of Fundamental Rights of European Union}.
\newblock
\newblock
\urldef\tempurl%
\url{https://www.europarl.europa.eu/charter/pdf/text_en.pdf}
\showURL{%
\tempurl}


\bibitem[{Council of European Union}(2004)]%
        {Dtravel}
\bibfield{author}{\bibinfo{person}{{Council of European Union}}.} \bibinfo{year}{2004}\natexlab{}.
\newblock \bibinfo{title}{Council Directive 2004/82/EC of 29 April 2004 on the obligation of carriers to communicate passenger data}.
\newblock
\newblock
\urldef\tempurl%
\url{http://data.europa.eu/eli/dir/2004/82/oj}
\showURL{%
\tempurl}


\bibitem[{Council of European Union}(2016a)]%
        {GDPR}
\bibfield{author}{\bibinfo{person}{{Council of European Union}}.} \bibinfo{year}{2016}\natexlab{a}.
\newblock \bibinfo{title}{General Data Protection Regulation, Regulation (EU) 2016/679}.
\newblock
\newblock
\urldef\tempurl%
\url{http://data.europa.eu/eli/reg/2016/679/oj}
\showURL{%
\tempurl}


\bibitem[{Council of European Union}(2016b)]%
        {LED}
\bibfield{author}{\bibinfo{person}{{Council of European Union}}.} \bibinfo{year}{2016}\natexlab{b}.
\newblock \bibinfo{title}{Law Enforcement Directive, Directive (EU) 2016/2016/680}.
\newblock
\newblock
\urldef\tempurl%
\url{http://data.europa.eu/eli/dir/2016/680/oj}
\showURL{%
\tempurl}


\bibitem[{Council of European Union}(2016c)]%
        {PNRD}
\bibfield{author}{\bibinfo{person}{{Council of European Union}}.} \bibinfo{year}{2016}\natexlab{c}.
\newblock \bibinfo{title}{Passenger Name Record Directive, Directive (EU) 2016/681}.
\newblock
\newblock
\urldef\tempurl%
\url{https://eur-lex.europa.eu/legal-content/EN/TXT/PDF/?uri=CELEX:32016L0681}
\showURL{%
\tempurl}


\bibitem[{Council of European Union}(2024)]%
        {AIA}
\bibfield{author}{\bibinfo{person}{{Council of European Union}}.} \bibinfo{year}{2024}\natexlab{}.
\newblock \bibinfo{title}{Artificial Intelligence Act, Regulation (EU) 2024/1689}.
\newblock
\newblock
\urldef\tempurl%
\url{http://data.europa.eu/eli/reg/2024/1689/oj}
\showURL{%
\tempurl}


\bibitem[Darling et~al\mbox{.}(2021)]%
        {darling2021you}
\bibfield{author}{\bibinfo{person}{John~A Darling}, \bibinfo{person}{Christopher~L Jerde}, {and} \bibinfo{person}{Adam~J Sepulveda}.} \bibinfo{year}{2021}\natexlab{}.
\newblock \showarticletitle{What do you mean by false positive?}
\newblock \bibinfo{journal}{\emph{Environmental DNA}} \bibinfo{volume}{3}, \bibinfo{number}{5} (\bibinfo{year}{2021}), \bibinfo{pages}{879--883}.
\newblock


\bibitem[De~Gregorio and Dunn(2022)]%
        {de2022european}
\bibfield{author}{\bibinfo{person}{Giovanni De~Gregorio} {and} \bibinfo{person}{Pietro Dunn}.} \bibinfo{year}{2022}\natexlab{}.
\newblock \showarticletitle{The European risk-based approaches: Connecting constitutional dots in the digital age}.
\newblock \bibinfo{journal}{\emph{Common Market Law Review}} \bibinfo{volume}{59}, \bibinfo{number}{2} (\bibinfo{year}{2022}).
\newblock


\bibitem[EDri(2019)]%
        {EDri}
\bibfield{author}{\bibinfo{person}{EDri}.} \bibinfo{year}{2019}\natexlab{}.
\newblock \bibinfo{booktitle}{\emph{Why EU passenger surveillance fails its purpose}}.
\newblock
\urldef\tempurl%
\url{https://edri.org/our-work/why-eu-passenger-surveillance-fails-its-purpose/}
\showURL{%
Retrieved 21 December 2025 from \tempurl}


\bibitem[Finnair(2024)]%
        {finnair2024privacypolicy}
\bibfield{author}{\bibinfo{person}{Finnair}.} \bibinfo{year}{2024}\natexlab{}.
\newblock \bibinfo{booktitle}{\emph{Finnair privacy policy}}.
\newblock
\urldef\tempurl%
\url{https://www.finnair.com/de-en/info/finnair-privacy-policy}
\showURL{%
Retrieved 19 November 2024 from \tempurl}


\bibitem[Frik et~al\mbox{.}(2023)]%
        {frik2023model}
\bibfield{author}{\bibinfo{person}{Alisa Frik}, \bibinfo{person}{Julia Bernd}, {and} \bibinfo{person}{Serge Egelman}.} \bibinfo{year}{2023}\natexlab{}.
\newblock \showarticletitle{A Model of Contextual Factors Affecting Older Adults’ Information-Sharing Decisions in the US}.
\newblock \bibinfo{journal}{\emph{ACM Transactions on Computer-Human Interaction}} \bibinfo{volume}{30}, \bibinfo{number}{1} (\bibinfo{year}{2023}), \bibinfo{pages}{1--48}.
\newblock


\bibitem[Gellert(2016)]%
        {gellert2016we}
\bibfield{author}{\bibinfo{person}{Raphael Gellert}.} \bibinfo{year}{2016}\natexlab{}.
\newblock \showarticletitle{We have always managed risks in data protection law: understanding the similarities and differences between the rights-based and the risk-based approaches to data protection}.
\newblock \bibinfo{journal}{\emph{Eur. Data Prot. L. Rev.}}  \bibinfo{volume}{2} (\bibinfo{year}{2016}), \bibinfo{pages}{481}.
\newblock


\bibitem[Guinchard(2018)]%
        {guinchard2018taking}
\bibfield{author}{\bibinfo{person}{Audrey Guinchard}.} \bibinfo{year}{2018}\natexlab{}.
\newblock \showarticletitle{Taking proportionality seriously: The use of contextual integrity for a more informed and transparent analysis in EU data protection law}.
\newblock \bibinfo{journal}{\emph{European Law Journal}} \bibinfo{volume}{24}, \bibinfo{number}{6} (\bibinfo{year}{2018}), \bibinfo{pages}{434--457}.
\newblock


\bibitem[Haggerty and Ericson(2017)]%
        {haggerty2017surveillant}
\bibfield{author}{\bibinfo{person}{Kevin~D Haggerty} {and} \bibinfo{person}{Richard~V Ericson}.} \bibinfo{year}{2017}\natexlab{}.
\newblock \showarticletitle{The surveillant assemblage}.
\newblock \bibinfo{journal}{\emph{Surveillance, crime and social control}} (\bibinfo{year}{2017}), \bibinfo{pages}{61--78}.
\newblock


\bibitem[Haitsma(2023)]%
        {haitsma2023regulating}
\bibfield{author}{\bibinfo{person}{Lucas~Michael Haitsma}.} \bibinfo{year}{2023}\natexlab{}.
\newblock \showarticletitle{Regulating algorithmic discrimination through adjudication: the Court of Justice of the European Union on discrimination in algorithmic profiling based on PNR data}.
\newblock \bibinfo{journal}{\emph{Frontiers in Political Science}}  \bibinfo{volume}{5} (\bibinfo{year}{2023}), \bibinfo{pages}{1232601}.
\newblock


\bibitem[Heise(2010)]%
        {heise2010surveying}
\bibfield{author}{\bibinfo{person}{David~R Heise}.} \bibinfo{year}{2010}\natexlab{}.
\newblock \bibinfo{booktitle}{\emph{Surveying cultures: Discovering shared conceptions and sentiments}}.
\newblock \bibinfo{publisher}{John Wiley \& Sons}.
\newblock


\bibitem[Herrmann et~al\mbox{.}(2016)]%
        {herrmann2016privacy}
\bibfield{author}{\bibinfo{person}{Michael Herrmann}, \bibinfo{person}{Mireille Hildebrandt}, \bibinfo{person}{Laura Tielemans}, {and} \bibinfo{person}{Claudia Diaz}.} \bibinfo{year}{2016}\natexlab{}.
\newblock \showarticletitle{Privacy in location-based services: An interdisciplinary approach}.
\newblock \bibinfo{journal}{\emph{SCRIPTed}}  \bibinfo{volume}{13} (\bibinfo{year}{2016}), \bibinfo{pages}{144}.
\newblock


\bibitem[Hildebrandt(2014)]%
        {hildebrandt2014location}
\bibfield{author}{\bibinfo{person}{Mireille Hildebrandt}.} \bibinfo{year}{2014}\natexlab{}.
\newblock \showarticletitle{Location Data, Purpose Binding and Contextual Integrity: What’s the Message?}
\newblock \bibinfo{journal}{\emph{Protection of Information and the Right to Privacy-A New Equilibrium?}} (\bibinfo{year}{2014}), \bibinfo{pages}{31--62}.
\newblock


\bibitem[International(2023)]%
        {transparency2023Corruption}
\bibfield{author}{\bibinfo{person}{Transparency International}.} \bibinfo{year}{2023}\natexlab{}.
\newblock \bibinfo{title}{2023 {C}orruption {P}erceptions {I}ndex - {E}xplore {F}inland’s results --- transparency.org}.
\newblock \bibinfo{howpublished}{\url{https://www.transparency.org/en/cpi/2023/index/fin}}.
\newblock


\bibitem[Jefferson(2018)]%
        {jefferson2018predictable}
\bibfield{author}{\bibinfo{person}{Brian~Jordan Jefferson}.} \bibinfo{year}{2018}\natexlab{}.
\newblock \showarticletitle{Predictable policing: Predictive crime mapping and geographies of policing and race}.
\newblock \bibinfo{journal}{\emph{Annals of the American Association of Geographers}} \bibinfo{volume}{108}, \bibinfo{number}{1} (\bibinfo{year}{2018}), \bibinfo{pages}{1--16}.
\newblock


\bibitem[Kitchin and Dodge(2014)]%
        {kitchin2014code}
\bibfield{author}{\bibinfo{person}{Rob Kitchin} {and} \bibinfo{person}{Martin Dodge}.} \bibinfo{year}{2014}\natexlab{}.
\newblock \bibinfo{booktitle}{\emph{Code/space: Software and everyday life}}.
\newblock \bibinfo{publisher}{MIT Press}.
\newblock


\bibitem[Koops(2021)]%
        {koops2021concept}
\bibfield{author}{\bibinfo{person}{Bert-Jaap Koops}.} \bibinfo{year}{2021}\natexlab{}.
\newblock \showarticletitle{The concept of function creep}.
\newblock \bibinfo{journal}{\emph{Law, Innovation and Technology}} \bibinfo{volume}{13}, \bibinfo{number}{1} (\bibinfo{year}{2021}), \bibinfo{pages}{29--56}.
\newblock


\bibitem[Korff(2021)]%
        {korff2021opinion}
\bibfield{author}{\bibinfo{person}{Douwe Korff}.} \bibinfo{year}{2021}\natexlab{}.
\newblock \showarticletitle{Opinion on the Broader and Core Issues Arising in the PNR Case Currently Before the CJEU (Case C-817/19)}.
\newblock \bibinfo{journal}{\emph{Available at SSRN 4436951}} (\bibinfo{year}{2021}).
\newblock


\bibitem[Korff(2023)]%
        {korff2023did}
\bibfield{author}{\bibinfo{person}{Douwe Korff}.} \bibinfo{year}{2023}\natexlab{}.
\newblock \showarticletitle{Did the PNR judgment address the core issues raised by mass surveillance?}
\newblock \bibinfo{journal}{\emph{European Law Journal}} \bibinfo{volume}{29}, \bibinfo{number}{1-2} (\bibinfo{year}{2023}), \bibinfo{pages}{223--236}.
\newblock


\bibitem[Korff and Georges(2015)]%
        {Korff2015}
\bibfield{author}{\bibinfo{person}{Douwe Korff} {and} \bibinfo{person}{M Georges}.} \bibinfo{year}{2015}\natexlab{}.
\newblock \showarticletitle{Passenger Name Records, data mining \& data protection: the need for strong safeguards}.
\newblock \bibinfo{journal}{\emph{Council of Europe, Directorate General Human Rights and Rule of Law, T-PD}}  \bibinfo{volume}{11} (\bibinfo{year}{2015}).
\newblock


\bibitem[Lum and Isaac(2016)]%
        {lum2016predict}
\bibfield{author}{\bibinfo{person}{Kristian Lum} {and} \bibinfo{person}{William Isaac}.} \bibinfo{year}{2016}\natexlab{}.
\newblock \showarticletitle{To predict and serve?}
\newblock \bibinfo{journal}{\emph{Significance}} \bibinfo{volume}{13}, \bibinfo{number}{5} (\bibinfo{year}{2016}), \bibinfo{pages}{14--19}.
\newblock


\bibitem[Lumley(2004)]%
        {lumley2004analysis}
\bibfield{author}{\bibinfo{person}{Thomas Lumley}.} \bibinfo{year}{2004}\natexlab{}.
\newblock \showarticletitle{Analysis of complex survey samples}.
\newblock \bibinfo{journal}{\emph{Journal of statistical software}}  \bibinfo{volume}{9} (\bibinfo{year}{2004}), \bibinfo{pages}{1--19}.
\newblock


\bibitem[Lumley and Scott(2013)]%
        {lumley2013two}
\bibfield{author}{\bibinfo{person}{Thomas Lumley} {and} \bibinfo{person}{Alastair~J Scott}.} \bibinfo{year}{2013}\natexlab{}.
\newblock \showarticletitle{Two-sample rank tests under complex sampling}.
\newblock \bibinfo{journal}{\emph{Biometrika}} \bibinfo{volume}{100}, \bibinfo{number}{4} (\bibinfo{year}{2013}), \bibinfo{pages}{831--842}.
\newblock


\bibitem[Medlar et~al\mbox{.}(2022)]%
        {medlar2022critiquing}
\bibfield{author}{\bibinfo{person}{Alan Medlar}, \bibinfo{person}{Jing Li}, \bibinfo{person}{Yang Liu}, {and} \bibinfo{person}{Dorota Glowacka}.} \bibinfo{year}{2022}\natexlab{}.
\newblock \showarticletitle{Critiquing-based Modeling of Subjective Preferences}. In \bibinfo{booktitle}{\emph{Proceedings of the 30th ACM Conference on User Modeling, Adaptation and Personalization}}. \bibinfo{pages}{234--242}.
\newblock


\bibitem[Medlar et~al\mbox{.}(2017)]%
        {medlar2017}
\bibfield{author}{\bibinfo{person}{Alan Medlar}, \bibinfo{person}{Joel Pyykk\"{o}}, {and} \bibinfo{person}{Dorota Glowacka}.} \bibinfo{year}{2017}\natexlab{}.
\newblock \showarticletitle{Towards Fine-Grained Adaptation of Exploration/Exploitation in Information Retrieval}. In \bibinfo{booktitle}{\emph{Proceedings of the 22nd International Conference on Intelligent User Interfaces}}. \bibinfo{pages}{623–627}.
\newblock


\bibitem[Munk(2017)]%
        {munk2017100}
\bibfield{author}{\bibinfo{person}{Timme~Bisgaard Munk}.} \bibinfo{year}{2017}\natexlab{}.
\newblock \showarticletitle{100,000 false positives for every real terrorist: Why anti-terror algorithms don't work}.
\newblock \bibinfo{journal}{\emph{First Monday}} (\bibinfo{year}{2017}).
\newblock


\bibitem[Nissenbaum(2009)]%
        {10.5555/1822585}
\bibfield{author}{\bibinfo{person}{Helen Nissenbaum}.} \bibinfo{year}{2009}\natexlab{}.
\newblock \bibinfo{booktitle}{\emph{Privacy in Context: Technology, Policy, and the Integrity of Social Life}}.
\newblock \bibinfo{publisher}{Stanford University Press}.
\newblock
\showISBNx{0804752370}


\bibitem[Nissenbaum(2019)]%
        {nissenbaum2019contextual}
\bibfield{author}{\bibinfo{person}{Helen Nissenbaum}.} \bibinfo{year}{2019}\natexlab{}.
\newblock \showarticletitle{Contextual integrity up and down the data food chain}.
\newblock \bibinfo{journal}{\emph{Theoretical inquiries in law}} \bibinfo{volume}{20}, \bibinfo{number}{1} (\bibinfo{year}{2019}), \bibinfo{pages}{221--256}.
\newblock


\bibitem[Nolte et~al\mbox{.}(2024)]%
        {Nolte2024}
\bibfield{author}{\bibinfo{person}{Henrik Nolte}, \bibinfo{person}{Miriam Rateike}, {and} \bibinfo{person}{Michèle Finck}.} \bibinfo{year}{2024}\natexlab{}.
\newblock \showarticletitle{Robustness and Cybersecurity in the EU Artificial Intelligence Act}. In \bibinfo{booktitle}{\emph{Generative AI and Law (GenLaw ’24) Workshop at 41 International Conference on Machine Learning, Vienna, Austria}}.
\newblock


\bibitem[OECD(2024)]%
        {OECDFinland}
\bibfield{author}{\bibinfo{person}{OECD}.} \bibinfo{year}{2024}\natexlab{}.
\newblock \bibinfo{title}{{O}{E}{C}{D} {S}urvey on {D}rivers of {T}rust in {P}ublic {I}nstitutions 2024 {R}esults - {C}ountry {N}otes: {F}inland}.
\newblock \bibinfo{howpublished}{\url{https://www.oecd.org/en/publications/oecd-survey-on-drivers-of-trust-in-public-institutions-2024-results-country-notes_a8004759-en/finland_596ba5da-en.html}}.
\newblock


\bibitem[Olsen and Wiesener(2021)]%
        {olsen2021beyond}
\bibfield{author}{\bibinfo{person}{Henrik~Palmer Olsen} {and} \bibinfo{person}{Cornelius Wiesener}.} \bibinfo{year}{2021}\natexlab{}.
\newblock \showarticletitle{Beyond data protection concerns--the European passenger name record system}.
\newblock \bibinfo{journal}{\emph{Law, Innovation and Technology}} \bibinfo{volume}{13}, \bibinfo{number}{2} (\bibinfo{year}{2021}), \bibinfo{pages}{398--421}.
\newblock


\bibitem[on~the Openness~of Government~Activities(1999)]%
        {OpennessAct}
\bibfield{author}{\bibinfo{person}{The Finnish~Act on~the Openness~of Government~Activities}.} \bibinfo{year}{1999}\natexlab{}.
\newblock
\newblock
\urldef\tempurl%
\url{https://www.finlex.fi/fi/laki/ajantasa/1999/19990621}
\showURL{%
\tempurl}


\bibitem[Oomen et~al\mbox{.}(2024)]%
        {oomen2023rethinking}
\bibfield{author}{\bibinfo{person}{Tessa Oomen}, \bibinfo{person}{Jo{\~a}o Gon{\c{c}}alves}, {and} \bibinfo{person}{Anouk Mols}.} \bibinfo{year}{2024}\natexlab{}.
\newblock \showarticletitle{Rethinking Artificial Intelligence: Algorithmic Bias and Ethical Issues| Rage Against the Artificial Intelligence? Understanding Contextuality of Algorithm Aversion and Appreciation}.
\newblock \bibinfo{journal}{\emph{International Journal of Communication}}  \bibinfo{volume}{18} (\bibinfo{year}{2024}), \bibinfo{pages}{25}.
\newblock


\bibitem[Orr{\`u}(2022)]%
        {orru2022european}
\bibfield{author}{\bibinfo{person}{Elisa Orr{\`u}}.} \bibinfo{year}{2022}\natexlab{}.
\newblock \showarticletitle{The European PNR Directive as an instance of pre-emptive, risk-based algorithmic security and its implications for the regulatory framework 1}.
\newblock \bibinfo{journal}{\emph{Information Polity}} \bibinfo{volume}{27}, \bibinfo{number}{2} (\bibinfo{year}{2022}), \bibinfo{pages}{131--146}.
\newblock


\bibitem[O’Neill(2022)]%
        {o2022contextual}
\bibfield{author}{\bibinfo{person}{Elizabeth O’Neill}.} \bibinfo{year}{2022}\natexlab{}.
\newblock \showarticletitle{Contextual integrity as a general conceptual tool for evaluating technological change}.
\newblock \bibinfo{journal}{\emph{Philosophy \& Technology}} \bibinfo{volume}{35}, \bibinfo{number}{3} (\bibinfo{year}{2022}), \bibinfo{pages}{79}.
\newblock


\bibitem[O’Neill(2023)]%
        {o2023balancing}
\bibfield{author}{\bibinfo{person}{Elizabeth O’Neill}.} \bibinfo{year}{2023}\natexlab{}.
\newblock \showarticletitle{Balancing Caution and the Need for Change: The General Contextual Integrity Approach}.
\newblock \bibinfo{journal}{\emph{Philosophy \& Technology}} \bibinfo{volume}{36}, \bibinfo{number}{4} (\bibinfo{year}{2023}), \bibinfo{pages}{68}.
\newblock


\bibitem[Palmiotto(2024)]%
        {palmiotto2024decision}
\bibfield{author}{\bibinfo{person}{Francesca Palmiotto}.} \bibinfo{year}{2024}\natexlab{}.
\newblock \showarticletitle{When Is a Decision Automated? A Taxonomy for a Fundamental Rights Analysis}.
\newblock \bibinfo{journal}{\emph{German Law Journal}} \bibinfo{volume}{25}, \bibinfo{number}{2} (\bibinfo{year}{2024}), \bibinfo{pages}{210--236}.
\newblock


\bibitem[Pr{\'e}gent(2025)]%
        {pregent2025you}
\bibfield{author}{\bibinfo{person}{Alexandra Pr{\'e}gent}.} \bibinfo{year}{2025}\natexlab{}.
\newblock \showarticletitle{Why you Should not use CI to Evaluate Socially Disruptive Technology}.
\newblock \bibinfo{journal}{\emph{Philosophy \& Technology}} \bibinfo{volume}{38}, \bibinfo{number}{1} (\bibinfo{year}{2025}), \bibinfo{pages}{1--19}.
\newblock


\bibitem[Ramesh et~al\mbox{.}(2022)]%
        {ramesh2022hierarchical}
\bibfield{author}{\bibinfo{person}{Aditya Ramesh}, \bibinfo{person}{Prafulla Dhariwal}, \bibinfo{person}{Alex Nichol}, \bibinfo{person}{Casey Chu}, {and} \bibinfo{person}{Mark Chen}.} \bibinfo{year}{2022}\natexlab{}.
\newblock \showarticletitle{Hierarchical text-conditional image generation with clip latents}.
\newblock \bibinfo{journal}{\emph{arXiv preprint arXiv:2204.06125}} (\bibinfo{year}{2022}).
\newblock


\bibitem[Rule(2019)]%
        {rule2019contextual}
\bibfield{author}{\bibinfo{person}{James~B Rule}.} \bibinfo{year}{2019}\natexlab{}.
\newblock \showarticletitle{Contextual integrity and its discontents: A critique of Helen Nissenbaum's normative arguments}.
\newblock \bibinfo{journal}{\emph{Policy \& Internet}} \bibinfo{volume}{11}, \bibinfo{number}{3} (\bibinfo{year}{2019}), \bibinfo{pages}{260--279}.
\newblock


\bibitem[Sanfilippo et~al\mbox{.}(2018)]%
        {sanfilippo2018privacy}
\bibfield{author}{\bibinfo{person}{Madelyn Sanfilippo}, \bibinfo{person}{Brett Frischmann}, {and} \bibinfo{person}{Katherine Standburg}.} \bibinfo{year}{2018}\natexlab{}.
\newblock \showarticletitle{Privacy as commons: Case evaluation through the governing knowledge commons framework}.
\newblock \bibinfo{journal}{\emph{Journal of Information Policy}}  \bibinfo{volume}{8} (\bibinfo{year}{2018}), \bibinfo{pages}{116--166}.
\newblock


\bibitem[Sanfilippo and Frischmann(2023)]%
        {sanfilippo2023slow}
\bibfield{author}{\bibinfo{person}{Madelyn~Rose Sanfilippo} {and} \bibinfo{person}{Brett Frischmann}.} \bibinfo{year}{2023}\natexlab{}.
\newblock \showarticletitle{Slow-governance in smart cities: An empirical study of smart intersection implementation in four US college towns}.
\newblock \bibinfo{journal}{\emph{Internet Policy Review}} \bibinfo{volume}{12}, \bibinfo{number}{1} (\bibinfo{year}{2023}).
\newblock


\bibitem[Sanfilippo et~al\mbox{.}(2020)]%
        {sanfilippo2020disaster}
\bibfield{author}{\bibinfo{person}{Madelyn~R Sanfilippo}, \bibinfo{person}{Yan Shvartzshnaider}, \bibinfo{person}{Irwin Reyes}, \bibinfo{person}{Helen Nissenbaum}, {and} \bibinfo{person}{Serge Egelman}.} \bibinfo{year}{2020}\natexlab{}.
\newblock \showarticletitle{Disaster privacy/privacy disaster}.
\newblock \bibinfo{journal}{\emph{Journal of the Association for Information Science and Technology}} \bibinfo{volume}{71}, \bibinfo{number}{9} (\bibinfo{year}{2020}), \bibinfo{pages}{1002--1014}.
\newblock


\bibitem[Selbst(2013)]%
        {selbst2013contextual}
\bibfield{author}{\bibinfo{person}{Andrew~D Selbst}.} \bibinfo{year}{2013}\natexlab{}.
\newblock \showarticletitle{Contextual expectations of privacy}.
\newblock \bibinfo{journal}{\emph{Cardozo L. Rev.}}  \bibinfo{volume}{35} (\bibinfo{year}{2013}), \bibinfo{pages}{643}.
\newblock


\bibitem[Shvartzshnaider et~al\mbox{.}(2019)]%
        {shvartzshnaider2019going}
\bibfield{author}{\bibinfo{person}{Yan Shvartzshnaider}, \bibinfo{person}{Noah Apthorpe}, \bibinfo{person}{Nick Feamster}, {and} \bibinfo{person}{Helen Nissenbaum}.} \bibinfo{year}{2019}\natexlab{}.
\newblock \showarticletitle{Going against the (appropriate) flow: A contextual integrity approach to privacy policy analysis}. In \bibinfo{booktitle}{\emph{Proceedings of the AAAI Conference on Human Computation and Crowdsourcing}}, Vol.~\bibinfo{volume}{7}. \bibinfo{pages}{162--170}.
\newblock


\bibitem[Sileno et~al\mbox{.}(2021)]%
        {sileno2021like}
\bibfield{author}{\bibinfo{person}{Giovanni Sileno}, \bibinfo{person}{Alexander Boer}, \bibinfo{person}{Geoff Gordon}, {and} \bibinfo{person}{Bernhard Rieder}.} \bibinfo{year}{2021}\natexlab{}.
\newblock \showarticletitle{Like circles in the water: Responsibility as a system-level function}. In \bibinfo{booktitle}{\emph{International Workshop on AI Approaches to the Complexity of Legal Systems}}. Springer, \bibinfo{pages}{198--211}.
\newblock


\bibitem[Smuha(2021)]%
        {smuha2021beyond}
\bibfield{author}{\bibinfo{person}{Nathalie~A Smuha}.} \bibinfo{year}{2021}\natexlab{}.
\newblock \showarticletitle{Beyond the individual: governing AI’s societal harm}.
\newblock \bibinfo{journal}{\emph{Internet Policy Review}} \bibinfo{volume}{10}, \bibinfo{number}{3} (\bibinfo{year}{2021}).
\newblock


\bibitem[Th{\"o}nnes(2022a)]%
        {thonnes2022cautious}
\bibfield{author}{\bibinfo{person}{Christian Th{\"o}nnes}.} \bibinfo{year}{2022}\natexlab{a}.
\newblock \showarticletitle{A cautious green light for technology-driven mass surveillance}.
\newblock \bibinfo{journal}{\emph{Verfassungsblog On Matters Constitutional}} (\bibinfo{year}{2022}).
\newblock


\bibitem[Th{\"o}nnes(2022b)]%
        {thonnes2022directive}
\bibfield{author}{\bibinfo{person}{Christian Th{\"o}nnes}.} \bibinfo{year}{2022}\natexlab{b}.
\newblock \showarticletitle{A directive altered beyond recognition: on the Court of Justice of the European Union’s PNR decision (C-817/19)}.
\newblock \bibinfo{journal}{\emph{Verfassungsblog On Matters Constitutional}} (\bibinfo{year}{2022}).
\newblock


\bibitem[Thönnes and Vavoula(2023)]%
        {Thönnes}
\bibfield{author}{\bibinfo{person}{Chrstian Thönnes} {and} \bibinfo{person}{Niovi Vavoula}.} \bibinfo{year}{2023}\natexlab{}.
\newblock \bibinfo{booktitle}{\emph{Automated predictive threat detection}}.
\newblock \bibinfo{publisher}{Max Planck Institute for the Study of Crime, Security and Law Working Paper}, Chapter~1.
\newblock


\bibitem[Trials(2021)]%
        {trials2021automating}
\bibfield{author}{\bibinfo{person}{Fair Trials}.} \bibinfo{year}{2021}\natexlab{}.
\newblock \bibinfo{title}{Automating injustice: The use of artificial intelligence \& automated decision-making systems in criminal justice in Europe}.
\newblock
\newblock


\bibitem[Tzanou(2010)]%
        {tzanou2010eu}
\bibfield{author}{\bibinfo{person}{Maria Tzanou}.} \bibinfo{year}{2010}\natexlab{}.
\newblock \showarticletitle{The EU as an emerging'Surveillance Society': The function creep case study and challenges to privacy and data protection}.
\newblock \bibinfo{journal}{\emph{ICL Journal}} \bibinfo{volume}{4}, \bibinfo{number}{3} (\bibinfo{year}{2010}), \bibinfo{pages}{407--427}.
\newblock


\bibitem[Ulbricht(2018)]%
        {ulbricht2018big}
\bibfield{author}{\bibinfo{person}{Lena Ulbricht}.} \bibinfo{year}{2018}\natexlab{}.
\newblock \showarticletitle{When big data meet securitization. Algorithmic regulation with passenger name records}.
\newblock \bibinfo{journal}{\emph{European Journal for Security Research}} \bibinfo{volume}{3}, \bibinfo{number}{2} (\bibinfo{year}{2018}), \bibinfo{pages}{139--161}.
\newblock


\bibitem[van~de Poel(2022)]%
        {van2022socially}
\bibfield{author}{\bibinfo{person}{Ibo van~de Poel}.} \bibinfo{year}{2022}\natexlab{}.
\newblock \showarticletitle{Socially disruptive technologies, contextual integrity, and conservatism about moral change}.
\newblock \bibinfo{journal}{\emph{Philosophy \& technology}} \bibinfo{volume}{35}, \bibinfo{number}{3} (\bibinfo{year}{2022}), \bibinfo{pages}{82}.
\newblock


\bibitem[Wernick and Artyushina(2023)]%
        {wernick2023future}
\bibfield{author}{\bibinfo{person}{Alina Wernick} {and} \bibinfo{person}{Anna Artyushina}.} \bibinfo{year}{2023}\natexlab{}.
\newblock \showarticletitle{Future-proofing the city: A human rights-based approach to governing algorithmic, biometric and smart city technologies}.
\newblock \bibinfo{journal}{\emph{Internet policy review}} \bibinfo{volume}{12}, \bibinfo{number}{1} (\bibinfo{year}{2023}), \bibinfo{pages}{1--26}.
\newblock


\bibitem[Williams and Kind(2019)]%
        {williams2019data}
\bibfield{author}{\bibinfo{person}{Patrick Williams} {and} \bibinfo{person}{Eric Kind}.} \bibinfo{year}{2019}\natexlab{}.
\newblock \bibinfo{title}{Data-driven Policing: The hardwiring of discriminatory policing practices across Europe.}
\newblock
\newblock


\bibitem[Wood et~al\mbox{.}(2006)]%
        {wood2006report}
\bibfield{author}{\bibinfo{person}{David~Murakami Wood}, \bibinfo{person}{Kirstie Ball}, \bibinfo{person}{David Lyon}, \bibinfo{person}{Clive Norris}, {and} \bibinfo{person}{Charles Raab}.} \bibinfo{year}{2006}\natexlab{}.
\newblock \showarticletitle{A report on the surveillance society}.
\newblock \bibinfo{journal}{\emph{Surveillance Studies Network, UK}} (\bibinfo{year}{2006}), \bibinfo{pages}{1--98}.
\newblock


\bibitem[Yeung(2019)]%
        {Yeungcoe}
\bibfield{author}{\bibinfo{person}{Karen Yeung}.} \bibinfo{year}{2019}\natexlab{}.
\newblock \bibinfo{title}{Responsibility and AI}.
\newblock
\newblock
\urldef\tempurl%
\url{https://rm.coe.int/responsability-and-ai-en/168097d9c5}
\showURL{%
\tempurl}


\bibitem[Zhang et~al\mbox{.}(2022)]%
        {zhang2022stop}
\bibfield{author}{\bibinfo{person}{Shikun Zhang}, \bibinfo{person}{Yan Shvartzshnaider}, \bibinfo{person}{Yuanyuan Feng}, \bibinfo{person}{Helen Nissenbaum}, {and} \bibinfo{person}{Norman Sadeh}.} \bibinfo{year}{2022}\natexlab{}.
\newblock \showarticletitle{Stop the spread: A contextual integrity perspective on the appropriateness of covid-19 vaccination certificates}. In \bibinfo{booktitle}{\emph{Proceedings of the 2022 ACM Conference on Fairness, Accountability, and Transparency}}. \bibinfo{pages}{1657--1670}.
\newblock


\bibitem[Zimmer(2018)]%
        {zimmer2018addressing}
\bibfield{author}{\bibinfo{person}{Michael Zimmer}.} \bibinfo{year}{2018}\natexlab{}.
\newblock \showarticletitle{Addressing conceptual gaps in big data research ethics: An application of contextual integrity}.
\newblock \bibinfo{journal}{\emph{Social Media+ Society}} \bibinfo{volume}{4}, \bibinfo{number}{2} (\bibinfo{year}{2018}).
\newblock


\end{thebibliography}

\clearpage
\appendix

\setcounter{table}{0}
\renewcommand{\thetable}{A\arabic{table}}

\section{Regression Models for Perceived Legitimacy of False Positives}

\begin{table}[h!]
\centering
\caption{Summary table for interval regression model of legitimate false positive count between air travel and sea travel controlling for gender. The baseline for \textit{travel mode} was air travel and the baseline for gender was \textit{female}. Coefficient estimates are log-transformed.}
\begin{tabular}{lllll}
\toprule
\multirow{2}{*}{} 
& \multirow{2}{*}{Estimate} 
& \multirow{2}{*}{\begin{tabular}[c]{@{}l@{}}Std.\\ Error\end{tabular}} 
& \multirow{2}{*}{$z$} 
& \multirow{2}{*}{$p$-value} \\ 
            &         &        &       &                 \\ \midrule
(Intercept) & 3.3816  & 0.0627 & 53.92 & \textless 2e-16$^{\ast\ast\ast}$ \\
travel mode  & -0.1650 & 0.0766 & -2.15 & 0.031$^{\ast}$           \\
gender      & -0.2528 & 0.0770 & -3.28 & 0.001$^{\ast\ast}$          \\ \bottomrule
\end{tabular}%
\label{tab:regression_travel_gender}
\end{table}

\begin{table}[h!]
\centering
\caption{Summary table for interval regression model of legitimate false positive count between air travel and sea travel. The baseline for \textit{travel mode} was air travel. Coefficient estimates are log-transformed.}
\begin{tabular}{lrrrl}
\toprule
\multirow{2}{*}{} 
& \multirow{2}{*}{Estimate} 
& \multirow{2}{*}{\begin{tabular}[c]{@{}l@{}}Std.\\ Error\end{tabular}} 
& \multirow{2}{*}{$z$} 
& \multirow{2}{*}{$p$-value} \\ 
    & & & & \\ \midrule
(Intercept)     & 3.2514    &  0.0523   & 62.20     & \textless 2e-16$^{\ast\ast\ast}$   \\
travel mode     & -0.1556   &  0.0769   & -2.02     & 0.043$^{\ast}$    \\ \bottomrule
\end{tabular}
\label{tab:regression_travel_only}
\end{table}

\section{Survey Questions}
\label{sec:appendix}

\begin{table*}
    \centering
    \footnotesize
    \caption{Demographic and background questions in English and Finnish. Square brackets indicate alternate wordings for air and sea conditions.}
    \renewcommand{\arraystretch}{1.2}
    \begin{tabular}{l p{0.4\textwidth} p{0.4\textwidth}} \toprule
        
        \multicolumn{3}{c}{Demographic and Background Information} \\ \midrule
        
        D1 & What is your gender? & Mikä on sukupuolesi? \\
        D2 & What is your age group? & Mihin ikäryhmään kuulut? \\
        D3 & What is your level of education? & Mikä on koulutustasosi? \\
        D4 & Would you describe yourself belonging to an ethnic group that is discriminated against in Finland? & 
        Kuvailisitko kuuluvasi etniseen ryhmään, jota syrjitään Suomessa? \\
        D5 & How many outbound trips do you take by [plane/ferry] in a typical year (excluding the pandemic)? & 
        Kuinka monta ulkomaanmatkaa teet [lentäen/laivalla] tyypillisenä vuonna (pois lukien pandemia-aika)? \\         
        \vspace{-0.5em} D6 & A majority of outbound trips I take by [plane/ferry] are for: \newline
        \hphantom{1} a) work \newline
        \hphantom{1} b) private reasons \newline
        \hphantom{1} c) I did not go on any trips \newline
        & 
        Suurin osa ulkomaanmatkoista, jotka teen [lentäen/laivalla] ovat: \newline
        \hphantom{1} a) työmatkoja \newline
        \hphantom{1} b) vapaa-ajan matkoja \newline
        \hphantom{1} c) en käynyt yhdelläkään matkalla \newline
        \\
        D7 & During the past year, I have been afraid of becoming a victim of a crime outside my home. & 
        Viimeisen vuoden aikana olen pelännyt joutuvani rikoksen uhriksi kotini ulkopuolella. \\
        D8 & During the past year, I have been afraid of becoming a victim of a terrorist attack. & 
        Viimeisen vuoden aikana olen pelännyt joutuvani terrori-iskun uhriksi. \\ \bottomrule

    \end{tabular}
    \label{tab:demographics_dual}
\end{table*}

\begin{table*}
    \centering
    \footnotesize
    \caption{Attitudinal questions on automated passenger pre-screening in English and Finnish. Square brackets indicate alternate wordings for air and sea conditions.}
    \renewcommand{\arraystretch}{1.2}
    \begin{tabular}{l p{0.4\textwidth} p{0.4\textwidth}} \toprule
        
        \multicolumn{3}{c}{Usage of passenger data} \\ \midrule
        
        Q1 & It is acceptable for the [airline/ferry company] to provide the police, customs and border guard with passenger data for the purpose of crime prevention. & On hyväksyttävää, että [lentoyhtiö/laivayhtiö] siirtää matkustajatiedot poliisille, Tullille ja Rajavartiolaitokselle käytettäväksi rikosten estämiseen. \\
        Q2 & It is acceptable for the [airline/ferry company] to provide Valvira (National supervisory authority for welfare and health) with passenger data for the purpose of preventing infectious disease transmission during a pandemic. & On hyväksyttävää, että [lentoyhtiö/laivayhtiö] siirtää matkustajatiedot Valviralle (Sosiaali- ja terveysalan lupa- ja valvontavirasto) käytettäväksi vaarallisten tartuntatautien leviämisen estämiseksi pandemiatilanteessa. \\
        Q3 & It is acceptable for the police, customs, and the border guard to use passenger data to automatically pre-screen passengers for potential terrorists and criminals. & On hyväksyttävää, että poliisi, Tulli ja Rajavartiolaitos käyttävät matkustajatietoja matkustajien automaattiseen seulontaan potentiaalisten terroristien ja rikollisen varalta. \\
        Q4 & In addition to passenger data, I would willingly give additional personal information (e.g., social media posts or credit card purchase information) to the police, customs, and the border guard, if it would make security checks faster and more convenient. & Matkustajatietojen lisäksi antaisin mielelläni enemmän henkilötietoja (esim. sosiaalisen median julkaisuja tai luottokortin ostotietoja) poliisille, Tullille ja Rajavartiolaitokselle, jos se tekisi turvatarkastuksista nopeampia ja miellyttävämpiä. \\
        Q5 & The use of my passenger data by the police, customs and the border guard makes me feel that the state does not trust me. & Matkustajatietojen käyttö poliisin, Tullin ja Rajavartiolaitoksen toimesta saa minut tuntemaan, että valtio ei luota minuun. \\ \midrule
        
        \multicolumn{3}{c}{Usage of automated computer analysis} \\ \midrule
        
        Q6 & It is acceptable to use passenger data for automated pre-screening of all passengers to identify potential criminal threats. & On hyväksyttävää käyttää matkustajatietoja kaikkien matkustajien automaattiseen seulontaan potentiaalisten rikosuhkien tunnistamiseksi. \\
        Q7 & I would rather my passenger data be pre-screened automatically by a computer than manually by a human. & Haluaisin ennemmin, että minun matkustajatietoni seuloo automatisoitu tietokone kuin ihminen manuaalisesti. \\
        Q8 & It is acceptable to flag passengers as suspicious if they exhibit travel patterns that resemble those of criminals (e.g., last minute bookings, payment in cash, unusually long or expensive routes, no baggage). & On hyväksyttävää merkitä matkustajia epäilyttäväksi, jos heidän matkustuskaavansa muistuttavat rikollisten matkustuskaavoja (esim. jokin näistä: viime hetken varaus, maksu käteisellä, epätavallisen pitkä tai kallis reitti, ei matkatavaroita). \\
        Q9 & Automated pre-screening of all passengers in [air/sea] travel based on passenger data is an excessive use of state power. & Kaikkien matkustajien automaattinen [ennakkoseulonta/ laivamatkustuksessa] matkustajatietojen perusteella on liiallista valtion vallankäyttöä. \\ \midrule
        
        \multicolumn{3}{c}{Consequences of automated computer analysis} \\ \midrule
        
        \vspace{-0.5em} Q10 & It is acceptable for police officers to manually examine the following information about individuals who have been flagged during passengers pre-screening: \newline
            \hphantom{1} a) Information on police databases (e.g., connections to ongoing criminal investigations) \newline
            \hphantom{1} b) Criminal record \newline
            \hphantom{1} c) Information in the Population Information System (including personal and family data, details of property and building ownership, and previous name(s) and address(es), current and previous memberships of a religious community) \newline
            \hphantom{1} d) MyKanta (containing health records and prescriptions) \newline
            \hphantom{1} e) Credit register (containing information on defaulted payments) \newline
            \hphantom{1} f) Social media \newline
            \hphantom{1} g) News about the person that can be found online \newline
            \hphantom{1} h) None of the above are acceptable \newline
        & 

        On hyväksyttävää, että poliisi tarkastaa manuaalisesti automaattisessa ennakkoseulonnassa merkittyä henkilöä koskevat tiedot: \newline
            \hphantom{1} a) Tiedot poliisin rekistereissä (esim. liitynnät käynnissä oleviin esitutkintoihin) \newline
            \hphantom{1} b) Rikosrekisteri \newline
            \hphantom{1} c) Väestörekisteritiedot (sisältää mm. henkilö- ja perhetiedot, kiinteistö- ja rakennusomistukset, aikaisemmat nimet ja osoitteet, uskonnollisen yhdyskunnan nykyisen tai aiemman jäsenyyden) \newline
            \hphantom{1} d) Omakanta (sis. terveystiedot ja reseptit) \newline
            \hphantom{1} e) Luottotietorekisterin (sis. maksuhäiriömerkinnät) \newline
            \hphantom{1} f) Sosiaalinen media \newline
            \hphantom{1} g) Internetistä löytyvät uutiset kyseisestä henkilöstä \newline
            \hphantom{1} h) Mikään näistä ei ole hyväksyttävää \newline
        \\
        Q11 & It is acceptable for police officers to stop flagged passengers at the [airport/ferry terminal] to confirm whether the person is an actual criminal threat. & On hyväksyttävää, että poliisi pysäyttää merkityn matkustajan [lentokentällä/laivaterminaalissa] varmistaakseen onko henkilö todellinen rikosuhka. \\
        Q12 & It is acceptable for police officers to stop flagged passengers at the [airport/ferry terminal] to confirm whether the person is an actual criminal threat even if it causes them to miss their [flight/departure]. & On hyväksyttävää, että poliisi pysäyttää merkityn matkustajan lentokentällä varmistaakseen onko henkilö todellinen rikosuhka, vaikka kyseinen henkilö myöhästyisi tämän takia [lennolta/laivasta]. \\ \bottomrule
    \end{tabular}
    \label{tab:attitudes_dual}
\end{table*}

\begin{table*}
    \centering
    \footnotesize
    \caption{False positive vignette in English and Finnish. Square brackets indicate alternate wordings for air and sea conditions. In this example, 4 passengers were flagged by passenger pre-screening: 1 true positive and 3 false positives.}
    \renewcommand{\arraystretch}{1.2}
    \begin{tabular}{l p{0.4\textwidth} p{0.4\textwidth}} \toprule
        
        Page & \multicolumn{2}{c}{False Positive Vignette} \\ \midrule
        
        \vspace{1em} \multirow{2}{*}{1}
        & The following image shows a typical outcome of using passenger data for automated pre-screening before the departure of an [airplane/ferry]: 4/200 passengers were flagged as suspicious. 
        & Seuraava kuva näyttää matkustajatietojen perusteella tehdyn automaattisen ennakkoseulonnan tyypillisen tuloksen ennen [lennon/matkustajalaivan] lähtöä: 4/200 matkustajaa merkittiin epäilyttäväksi. \\

        \vspace{1em} & \multicolumn{2}{c}{ \includegraphics[width=0.6\textwidth]{manifest-4-200.png} } \\
        
        \vspace{1em} \multirow{2}{*}{2}
        & These 1/4 passengers were prevented from boarding the [plane/ferry] by the police who determined they presented a genuine criminal threat after a thorough examination of their personal information. 
        & Näiden 1/4 matkustajan pääsy [lennolle/laivalle] estettiin poliisin toimesta, koska matkustajien todettiin olevan todellinen rikosuhka sen jälkeen, kun heidän henkilötietonsa oli perusteellisesti tutkittu. \\

        \vspace{1em} & \multicolumn{2}{c}{ \includegraphics[width=0.166\textwidth]{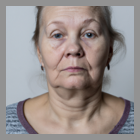} } \\
        
        \vspace{1em} \multirow{2}{*}{3}
        & These 3/4 passengers were re-classified manually as false alarms by the police after their personal information had been thoroughly examined. These passengers boarded the [plane/ferry], but were unaware that they had been flagged by the surveillance system and that the police had examined their personal information.
        & Nämä 3/4 matkustajaa uudelleenarvioitiin manuaalisesti vääriksi hälytyksiksi poliisin toimesta sen jälkeen, kun heidän henkilötietonsa oli perusteellisesti tutkittu. Nämä matkustajat pääsivät [lennolle/laivalle] tietämättä siitä, että tietojärjestelmä oli merkinnyt heidät ja poliisi oli tutkinut heidän tietonsa. \\

        \vspace{1em} & \multicolumn{2}{c}{ \includegraphics[width=0.5\textwidth]{face_fp.png} } \\
        
        \multirow{2}{*}{4}
        & Consider the following moral dilemma: 
Automated pre-screening is not foolproof and produces false alarms. When the surveillance system flags a passenger, it always leads to detailed examination of their personal information even though they may be innocent. 
        & Mitä ajattelet seuraavasta moraalisesta pulmasta:
Automaattinen ennakkoseulonta ei ole vedenpitävää ja tuottaa vääriä hälytyksiä. Kun tietojärjestelmä antaa hälytyksen, se johtaa aina merkityn henkilön yksityiskohtaiseen tarkasteluun siitä huolimatta, että merkitty henkilö saattaa olla viaton matkustaja. \\

        & Given the number of false alarms presented above, which statement are you more in agreement with: \newline
            \hphantom{1} a) I prioritise identifying criminal threats over the privacy of innocent passengers, even if it means more false alarms and privacy violations of innocent passengers by the police. \newline
            \hphantom{1} b) I prioritise protecting innocent passengers’ privacy, even if reducing false alarms also means fewer alerts of criminal threats. \newline
        
        & Ottaen huomioon yllä esitettyjen väärien hälytysten määrä, kumman väittämän kanssa olet enemmän samaa mieltä: \newline
            \hphantom{1} a) Priorisoin ennemmin todellisten rikosuhkien tunnistamista kuin viattomien matkustajien yksityisyyttä, vaikka se tarkoittaisi enemmän vääriä hälytyksiä ja viattomien matkustajien joutumista poliisin tarkan manuaalisen tarkastelun kohteeksi. \newline
            \hphantom{1} b) Priorisoin ennemmin viattomien matkustajien yksityisyyden suojelua poliisin tarkalta manuaaliselta tutkinnalta, vaikka väärien hälytysten vähentäminen tarkoittaisi, että järjestelmä antaa vähemmän hälytyksiä todellisista rikosuhkista. \newline
        
        \\ \bottomrule

    \end{tabular}
    \label{tab:vignette_dual}
\end{table*}

\end{document}